\documentclass[aps,prd,floats,floatfix,twocolumn,nofootinbib,showpacs]{revtex4-1}

\usepackage{verbatim}
\usepackage{amsmath}
\usepackage{amsfonts}
\usepackage{amssymb}
\usepackage{mathrsfs}
\usepackage{hyperref}
\usepackage{rotating}
\usepackage{upgreek}
\usepackage{bm}
\usepackage{dcolumn}
\usepackage{epsfig}
\usepackage{graphicx}
\usepackage{graphics}
\usepackage[latin1]{inputenc}
\usepackage{latexsym}
\usepackage[usenames]{color}
\usepackage{yfonts}

\usepackage{graphicx}
\usepackage{xspace} 
\usepackage[usenames]{color}
\usepackage{ulem}
\definecolor {darkgreen}{rgb}{0.2,0.7,0.2}

\newcommand{\be}{\begin{equation}}
\newcommand{\ee}{\end{equation}}
\newcommand{\ben}{\begin{equation*}}
\newcommand{\een}{\end{equation*}}
\newcommand{\bea}{\begin{eqnarray}}
\newcommand{\eea}{\end{eqnarray}}
\newcommand{\ba}{\begin{eqnarray}}
\newcommand{\ea}{\end{eqnarray}}
\newcommand{\bse}{\begin{subequations}}
\newcommand{\ese}{\end{subequations}}
\newcommand{\bal}{\begin{align}}
\newcommand{\eal}{\end{align}}
\newcommand{\nn}{\nonumber}
\newcommand{\bml}{\begin{multline}}
\newcommand{\eml}{\end{multline}}
\newcommand{\bean}{\begin{eqnarray*}}
\newcommand{\eean}{\end{eqnarray*}}
\newcommand{\pd}{\partial}
\newcommand{\cd}{\nabla}
\newcommand{\bcd}{\overline{\nabla}}
\newcommand{\rmd}{\mathrm{d}}

\newcommand{\<}{\langle}
\renewcommand{\>}{\rangle}
\newcommand{\scL}{\mathcal{L}}

\newcommand{\scO}{\mathcal{O}}

\newcommand{\scR}{\mathcal{R}}
\newcommand{\scri}{\mathscr{I}}
\newcommand{\barh}{\underline{\tilde{h}}}
\newcommand{\tildetheta}{\tilde{\vartheta}}

\newcommand{\GW}{{\mbox{\tiny GW}}}
\newcommand{\eff}{{\mbox{\tiny eff}}}
\newcommand{\INT}{{\mbox{\tiny int}}}
\newcommand{\DYN}{{\mbox{\tiny dyn}}}
\newcommand{\GR}{{\mbox{\tiny GR}}}
\newcommand{\EH}{{\mbox{\tiny EH}}}
\newcommand{\MT}{{\mbox{\tiny MT}}}

\newcommand{\TT}{{\mbox{\tiny TT}}}
\newcommand{\MAT}{{\mbox{\tiny mat}}}

\newcommand{\CS}{{\mbox{\tiny CS}}}

\newcommand{\starR}{{{}^\ast\!}R}
\newcommand{\pont}{\,\starR\,R}
\newcommand{\met}{\mbox{g}}

\DeclareMathOperator{\sign}{sign}

\begin{document}

  \title{Effective Gravitational Wave Stress-energy Tensor in Alternative Theories of Gravity}
  \author{Leo C. Stein}
  \affiliation{Department of Physics and MIT Kavli Institute, 77
    Massachusetts Avenue, Cambridge, MA 02139}
 \author{Nicol\'as Yunes}
  \affiliation{Department of Physics and MIT Kavli Institute, 77
    Massachusetts Avenue, Cambridge, MA 02139}

\begin{abstract}
The inspiral of binary systems in vacuum is controlled by the stress-energy of gravitational radiation and any other propagating degrees of freedom.
For gravitational waves, the dominant contribution is characterized by
an effective stress-energy tensor at future null infinity.
We employ perturbation theory and the short-wavelength approximation to compute this stress-energy tensor in a wide class of alternative theories. 
We find that this tensor is generally a modification of that first computed by Isaacson, 
where the corrections can dominate over the general relativistic term. 
In a wide class of theories, however, these corrections identically
vanish at asymptotically flat, future, null infinity, reducing the stress-energy tensor to Isaacson's.  
We exemplify this phenomenon by first considering dynamical Chern-Simons modified gravity, 
which corrects the action via a scalar field and the contraction of the Riemann tensor and
its dual. 
We then consider a wide class of theories with dynamical scalar fields coupled to higher-order curvature invariants, and show that the gravitational wave stress-energy tensor still reduces to Isaacson's.
The calculations presented in this paper are crucial to perform systematic tests of such
modified gravity theories through the orbital decay of binary pulsars or through gravitational
wave observations.
\end{abstract}

\date{\today \hspace{0.2truecm}}
\maketitle

\section{Introduction}

Feynman has argued that no matter how beautiful or elegant a certain theory is, or how authoritative its proponents, if it does not agree with experiments, then it must be wrong. For the past 40 years, this philosophy has been applied to gravitational theories with great success. Many modified gravity theories that were prominent in the 1970's, have now been essentially discarded, as they were found to disagree with Solar System experiments or binary pulsar observations~\cite{lrr-2006-3}. Similarly, this decade is beginning to bring a wealth of astrophysical information that will be used to constrain new modified gravity theories. In fact, precision double binary pulsar observations~\cite{Burgay:2003jj,Lyne:2004cj,Kramer:2006nb} have already allowed us to constrain modified theories to exciting new levels~\cite{Yunes:2008ua,2010PhRvD..82h2002Y}. Future gravitational wave (GW) observations on Earth, with the Advanced Laser Interferometer Gravitational Observatory (aLIGO)~\cite{ligo,Abramovici:1992ah,:2007kva}, aVIRGO~\cite{virgo} and its collaborators, and in space, through the Laser Interferometer Space Antenna (LISA)~\cite{lisa,Prince:2003aa,Danzmann:2003tv,Danzmann:2003ad}, will allow new precision tests of strong field gravity~\cite{Schutz:2009tz}.

Such tests of alternative theories of gravity will be very sensitive to the motion of compact bodies in a regime of spacetime where gravitational fields and velocities are large, i.e.~the so-called strong-field. Gralla~\cite{Gralla:2010cd} has shown that motion in classical field theories that satisfy certain conditions (the existence of a Bianchi-like identity and field equations no higher than second-order) is ``universally'' geodesic to leading-order in the binary system's mass-ratio, with possible deviations from geodesicity due to the bodies' internal structure. He also argues that one might be able to relax the second condition, as it does not seem necessary. In fact, motion in certain higher-order theories, such as Chern-Simons modified gravity~\cite{Alexander:2009tp}, is already known to be purely geodesic to leading-order in the mass-ratio, without influence of internal structure due to additional symmetries in the theory.  

Tests of modified gravity theories in the strong-field, however, not only require a prescription for the conservative sector of motion, but also of the dissipative sector, that which describes how the objects inspiral. Geodesic motion must thus be naturally corrected by a radiation-reaction force that drives non-geodesic motion toward an ultimate plunge and merger~\cite{Barack:2009ux}. Similarly, one can think of such motion as geodesic, but with {\emph{varying}} orbital elements~\cite{Pound:2007th,Gralla:2008fg,Pound:2009sm,Pound:2010pj,Pound:2010wa} (energy, angular momentum and Carter constant). The rate of change of such orbital elements is governed by the rate at which all degrees of freedom (gravitational and non-gravitational) radiate. 

In the gravitational sector and to leading order in the metric
perturbation, such a rate of change is controlled by an effective
stress-energy tensor for GWs, first computed by Isaacson in General
Relativity (GR)~\cite{Isaacson:1968ra,Isaacson:1968gw}. In his
approach, Isaacson expanded the Einstein equations to second order in
the metric perturbation about an arbitrary background. The first-order
equations describe the evolution of gravitational radiation. The
second-order equation serves as a source to the zeroth-order field
equations, just like a stress-energy tensor, and it depends on the
square of the first-order perturbation. This tensor can then be
averaged over several gravitational wavelengths, assuming the
background length scale is much longer than the GW wavelength (the
short-wavelength approximation). In this approximation, Isaacson found
that the effective GW energy-momentum tensor is proportional to the
square of  first partial derivatives of the metric perturbation,
i.e.~proportional to the square of the gravitational
frequency. Components of this stress-energy then provide the rate of
change of orbital elements, leading to the well-known quadrupole
formula.

Alternative theories of gravity generically lead to a modified effective GW stress-energy tensor. It is sometimes assumed that this stress-energy tensor will take the same form as in GR~\cite{Nelson:2010rt,Nelson:2010ru}, but this need not be the case. In GR, the scaling of this tensor with the GW frequency squared can be traced to the Einstein-Hilbert action's dependence on second-derivatives of the metric perturbation through the Ricci scalar. If the action is modified through the introduction of higher-powers of the curvature tensor, then the stress-energy tensor will be proportional to higher powers of the frequency. Therefore, the consistent calculation of the modified Isaacson tensor needs to be carried out until terms similar to the GR contribution (proportional to frequency squared) are obtained. This in turn implies that calculations of effective energy-momentum tensors in modified gravity theories to {\emph{leading-order in the GW frequency}} can sometimes be insufficient for determining the rate of change of orbital elements.

In this paper, we present a formalism to compute the energy-momentum
tensor consistently in generic classical field theories. We employ a
scheme where the action itself is first expanded in the metric
perturbation to second order, and the background metric and metric
perturbation are treated as independent fields.  Varying with respect
to the background metric leads to an effective GW stress-energy tensor
that can then be averaged over several wavelengths. This produces
results equivalent to Isaacson's calculation.

We exemplify this formulation by first considering CS gravity~\cite{Alexander:2009tp}. This theory modifies the Einstein-Hilbert action through the addition of the product of a scalar field with the contraction of the Riemann tensor and its dual. This scalar field is also given dynamics through a kinetic term in the action. The leading-order contribution to the CS-modified GW stress-energy tensor should appear at order frequency to the fourth-power, but Sopuerta and Yunes~\cite{Sopuerta:2009iy} have shown that this contribution vanishes at future null infinity.

We here continue this calculation through order frequency cubed and frequency squared and find that such CS modifications still vanish at future null infinity. This is because the background scalar field must decay at a certain rate for it to have a finite amount of energy in an asymptotically-flat spacetime. If one insists on ignoring such a requirement, such as in the case of the non-dynamical theory, then frequency-cubed CS modifications to the energy-momentum do not vanish. 

We explicitly calculate such modifications for a canonical embedding,
where the scalar field is a linear function of time in inertial
coordinates. This is similar to previous work~\cite{Guarrera:2007tu} that calculated another
effective stress-energy tensor for the non-dynamical version of Chern-Simons.
In this case, the dominant modification to the radiation-reaction force is in the rate of change of radiated momentum, which leads to so-called recoil velocities after binary coalescence. In GR, such recoil is proportional to the product of the (mass) quadrupole and octopole when multipolarly decomposing the radiation field. In non-dynamical CS gravity with a canonical embedding, the recoil is proportional to the square of the mass quadrupole, which dominates over the GR term. 

We then construct a wide class of alternative theories that
differ from GR through higher order curvature terms in the action
coupled to a scalar field. We
compute the GW stress-energy-momentum tensor in
such theories and find that corrections to the Isaacson tensor vanish
at future null infinity provided the following conditions are
satisfied: (i) the curvature invariants in the modification are quadratic
or higher order; (ii) the non-minimally coupled scalar field is dynamical; (iii)
the modification may be modeled as a weak deformation away from GR; (iv)
the spacetime is asymptotically flat at future null infinity. These results prove that the effective stress-energy tensor assumed in~\cite{Nelson:2010rt,Nelson:2010ru} is indeed correct\footnote{The authors of~\cite{Nelson:2010rt,Nelson:2010ru} presented an energy loss formula which was not evaluated at $\scri^+$. In the limit of $r\to\infty$, their energy loss formula reduces to the Isaacson formula.}.

Even if the effective GW energy-momentum tensor is identical to that
in GR, in terms of contractions of first derivatives of the metric
perturbation, this does not imply that GWs will not be
modified. First, background solutions could be modified. For example,
in dynamical CS gravity, the Kerr metric is not a solution to the
modified field equations for a rotating black hole
(BH)~\cite{Grumiller:2007rv}, but it is instead modified in the shift
sector~\cite{2009PhRvD..79h4043Y}. Second, the solution to the GW
evolution equation could also be modified. For example, in
non-dynamical CS gravity, GWs become amplitude birefringent as they
propagate~\cite{jackiw:2003:cmo,Alexander:2007kv,Yunes:2010yf}.
Third, additional degrees of freedom may also be present and radiate,
thus changing the orbital evolution.
All of these facts imply that even if the Isaacson tensor correctly
describes the effective GW energy-momentum tensor, GWs themselves can
and generically will be modified in such alternative theories.

In the remainder of this paper we use the following conventions. Background quantities are always denoted with an overhead bar, while perturbed quantities of first-order with an overhead tilde. 
We employ decompositions of the type 
$\met_{\mu \nu} = \bar{\met}_{\mu \nu} + \epsilon \; \tilde{h}_{\mu \nu} +
\scO(\epsilon^2)$, where $\met_{\mu \nu}$ is the full metric, $\bar{\met}_{\mu \nu}$ is the background metric and $\tilde{h}_{\mu \nu}$ is a small perturbation ($\epsilon \ll 1$ is a book-keeping parameter). Covariant differentiation with respect to the background metric is denoted via $\bar\nabla^{}_\mu B^{}_{\nu}$, while covariant differentiation with respect to the full metric is denoted via $\nabla^{}_\mu B^{}_{\nu}$. Symmetrization and antisymmetrization are denoted with parentheses and square brackets around the indices respectively, such as $A^{}_{(\mu\nu)}\equiv [A^{}_{\mu\nu} + A^{}_{\nu\mu}]/2$ and $A^{}_{[\mu\nu]} \equiv [A^{}_{\mu\nu} - A^{}_{\nu\mu}]/2$. We use the metric signature $(-,+,+,+)$ and geometric units, such that $G = c = 1$.

This paper is organized as follows:
Section~\ref{sec:pert-Lag} describes the perturbed Lagrangian approach used in this paper to compute the effective GW stress-energy tensor. 
Section~\ref{sec:GRTab} applies this framework to GR. 
Section~\ref{sec:CS} discusses dynamical CS gravity.
Section~\ref{sec:CSTab} computes the full effective stress-energy tensor in this theory. 
Section~\ref{sec:gen-alt-theories} generalizes the calculation to a wider
class of alternative theories.
Section~\ref{sec:conc} concludes and points to future research. 

\section{Perturbed Lagrangian approach}
\label{sec:pert-Lag}

Isaacson~\cite{Isaacson:1968ra,Isaacson:1968gw} introduced what is now the 
standard technique to obtain an effective stress-energy
tensor for gravitational radiation, via second-order perturbation theory 
on the equations of motion. This technique requires an averaging procedure 
to construct an \emph{effective} stress-energy tensor. This is because of 
the inability to localize the energy of the gravitational field to less than several 
wavelengths of the radiation, and because of the ambiguities of the
metric perturbation on distances of order the wavelength due to gauge
freedom\footnote{Gauge freedom in perturbation theory stands for the freedom to  
identify points between the physical and ``background'' manifolds~\cite{2005PThPh.113..481N,2007arXiv0711.0996N}.}. Isaacson employed the
Brill-Hartle~\cite{1964PhRv..135..271B} averaging scheme, although 
one can arrive at an identical quantity by using different schemes~\cite{1992GReGr..24.1015Z,1996GReGr..28..953Z,2004gr.qc....11004Z}, e.g.~Whitham or macroscopic gravity.

An alternative approach to derive field equations and an effective stress-energy tensor for GWs is to work at the level of the action. One possibility is to use the Palatini framework~\cite{Misner:1973cw,Carroll}, where the connection is promoted to an independent field that is varied in the action, together with the metric tensor. Such a framework, however is problematic in alternative theories of gravity, as it need not lead to the same field equations as variation of the action with respect to the metric tensor only.  

A similar but more appropriate approach is that of {\emph{second-variation}}~\cite{1973CMaPh..30..153M}. In this approach, the action is first expanded to second order in the metric perturbation, assuming the connection is the Christoffel one $\Gamma^\sigma_{\mu\nu}$. Then, the action is promoted to an effective one, by treating the background metric tensor and the metric perturbation as {\emph{independent fields}}. Variation of this effective action with respect to the metric perturbation and the background metric yields the equations of motion. The former variation leads to the first-order field equations, when the background field equations are imposed. The latter variation leads to the background field equations to zeroth order in the metric perturbation and to an effective GW stress-energy tensor to second order.  

Let us begin by expanding all quantities in a power series about a
background solution
\begin{equation}
\varphi = \bar{\varphi} + \epsilon \tilde{\varphi} +
\epsilon^2 \tilde{\tilde{\varphi}} + \scO(\epsilon^3)\ ,
\end{equation}
where $\epsilon \ll 1$ is an order counting parameter and 
$\varphi$ represents all tensor fields of the theory with
indices suppressed: $\bar\varphi$ is
the background field, $\tilde{\varphi}$ is the first-order perturbation to $\varphi$,
and $\tilde{\tilde{\varphi}}$ is the second-order
perturbation. The action can then be expanded, as
\begin{equation}
S[\varphi] = S^{(0)}[\bar\varphi] + 
S^{(1)}[\bar\varphi,\tilde{\varphi}] + 
S^{(2)}[\bar\varphi,\tilde{\varphi},\tilde{\tilde{\varphi}}] +
\scO(\epsilon^3) \,,
\label{schematic-exp}
\end{equation}
where $S^{(1)}$ is linear in $\tilde{\varphi}$ and $S^{(2)}$ is
quadratic in $\tilde{\varphi}$ but linear in $\tilde{\tilde{\varphi}}$.
We now define the effective action as Eq.~\eqref{schematic-exp} but
promoting $\bar{\varphi}$ and $\tilde{\varphi}$ to independent fields.

One might wonder why the field $\tilde{\tilde{\varphi}}$ is not also treated
as independent. First, variation of the action with respect to 
$\tilde{\tilde{\varphi}}$ would lead to second-order equations of motion, which
we are not interested in here. 
Second, the variation of the action with respect to $\tilde{\varphi}$ cannot introduce 
terms that depend on $\tilde{\tilde{\varphi}}$, because the product of $\tilde{\varphi}$ and 
$\tilde{\tilde{\varphi}}$ never appears in Eq.~\eqref{schematic-exp}, as this would be of 
${\cal{O}}(\epsilon^{3})$. Third, the variation of the action with respect to $\bar{\varphi}$ can
only introduce terms linear in $\tilde{\tilde{\varphi}}$, which vanish upon averaging, as
we describe in Sec.~\ref{sec:averageProps}. This is because averages of any odd number of short-wavelength quantities generically vanish. Therefore, we can safely neglect all terms that depend on $\tilde{\tilde{\varphi}}$ in the effective action, which renders Eq.~\eqref{schematic-exp} a functional of only $\bar{\varphi}$ and $\tilde{\varphi}$.

As in the standard approach, the second-order variation method still
requires that one performs a short-wavelength average of the effective
stress-energy tensor. Upon averaging, the variation of the first-order piece 
of the action $S^{(1)}$ with respect to $\bar{\varphi}$ vanishes because it generates
terms linear in $\tilde{\varphi}$.  Since $S^{(1)}$ does not contribute to the effective stress-energy tensor, we can safely drop it from the effective action for now. 

The effective action reduces to 
\begin{equation}
\label{eq:Seffgeneral}
S^\eff[\bar{\varphi},\tilde{\varphi}] = S^{\eff(0)}[\bar{\varphi}] +
S^{\eff(2)}[\bar{\varphi},\tilde{\varphi}] \ .
\end{equation}
Naturally, the variation of $S^{\eff(0)}$ with respect to the
background metric $\bar{\met}^{\mu\nu}$ yields the background
equations of motion. The effective stress-energy tensor comes from
averaging the variation of $S^{\eff(2)}$ with respect to
$\bar{\met}^{\mu\nu}$,
\begin{subequations}
\label{eq:Teffgeneral}
\begin{eqnarray}
\label{eq:TeffgeneralAction}
\delta S^{\eff(2)} &=& \epsilon^2 \int\rmd^4 x \sqrt{-\bar\met}\ \delta\bar\met^{\mu\nu}
t_{\mu\nu} \ , \\
\label{eq:TeffgeneralDef}
T^\eff_{\mu\nu} &\equiv& -2\epsilon^2\<\< t_{\mu\nu} \>\> \ ,
\end{eqnarray}
\end{subequations}
where the factor of $2$ is conventional for agreement with the
canonical stress-energy tensor, and $\<\< \ \>\>$ is the averaging
operator, which we discuss below. One of the immediate benefits of
working from an action principle comes from the diffeomorphism
invariance of the action. The diffeomorphism invariance immediately
implies that the variation of the total action with respect to the
metric is divergence free~\cite{Carroll}. When the matter
stress-energy tensor is itself divergence free, then the gravity
sector -- the sum of the stress-energy of non-minimally coupled
degrees of freedom and the effective stress-energy tensor of
gravitational waves -- will also be divergence free.

\subsection{Short-wavelength averaging}
\label{sec:averageProps} 

The goal of the averaging scheme is to distinguish radiative
quantities, those which are rapidly-varying functions of spacetime, from Coulomb-like
quantities, those which are slowly-varying functions. This is accomplished by defining the
operator $\<\<\ \>\>$ as a linear integral operator. This operator may either be an
average over the phase of the rapidly varying quantities or over
spacetime. If the integral is over spacetime, there is an averaging
kernel with characteristic length scale $L_{\mathrm{ave}}$ that
separates the foreground short-wavelength $\lambda_\GW$ from the
background length scale $L_{\mathrm{bg}}$, that is, $\lambda_\GW \ll
L_{\mathrm{ave}} \ll L_{\mathrm{bg}}$.

The details of the averaging scheme are not as important as their properties~\cite{1992GReGr..24.1015Z,1996GReGr..28..953Z,2004gr.qc....11004Z}, since one arrives at equivalent
results using different schemes. The most useful properties of $\<\<\ \>\>$ are:
\begin{enumerate}
\item The average of a product of an odd number of short-wavelength
  quantities vanishes.
\item The average of a derivative of a tensor vanishes, e.g. for some
  tensor expression $T^\mu{}_{\alpha\beta}$, $\<\< \bcd_\mu
  T^\mu{}_{\alpha\beta} \>\> = 0$.
\item As a corollary to the above, integration by parts can be
  performed, e.g. for tensor expressions
  $R^\mu{}_\alpha, S_\beta$, $\<\< R^\mu{}_\alpha \bcd_\mu S_\beta
  \>\> = - \<\< S_\beta \bcd_\mu R^\mu{}_\alpha \>\>$.
\end{enumerate}

Let us briefly mention where some of these properties come from and
some caveats. When considering monochromatic functions, an odd
oscillatory integral has an average value about zero, while an even
oscillatory integral has a non-zero average. This is enough to find
that that averages of expressions linear in a short-wavelength
quantity will vanish. Expressions at third (and higher odd) order
would vanish for monochromatic radiation, but not in general. However,
these are at sufficiently high order that we neglect them.

The vanishing of averages of derivatives is a subtle point. In
the spacetime average approach, this is found by integrating by parts,
leaving a term with a derivative on the averaging kernel. This term is
smaller than nonvanishing averages by a factor of
$\scO(\lambda_\GW/L_\mathrm{ave})$ and depends on the averaging
kernel. From physical grounds, the choice of averaging kernel should
not affect any physical quantities, so the average should in fact
vanish identically. From the action standpoint, the average of a
derivative can be seen to arise from an action term which is a total
divergence. Since total divergences in the action do not affect the
equations of motion, the average of a derivative vanishes.

A similar argument holds for integration by parts. In the Brill-Hartle
average scheme, integration by parts incurs an error of order
$\scO(\lambda_\GW/L_\mathrm{ave})$ from a derivative of the averaging
kernel. From the action standpoint, though, integration by parts at
the level of the action incurs no error, since there is no averaging
kernel in the action.

The fact that the variation of $S^{(1)}$ with respect to $\bar{\varphi}$ does not contribute 
to the effective stress-energy tensor is a direct consequence of property
(1) above. The vanishing of all terms linear in $\tilde{\tilde{\varphi}}$ upon averaging
is also a consequence of (1). As one can see, these properties greatly simplify all further calculations.

\subsection{Varying Christoffel and curvature tensors}
\label{sec:varR}

Let us now consider what types of terms arise from the variation
of the effective action with respect to $\bar{\met}^{\mu\nu}$. In order
to perform this variation properly, any implicit dependence of the action
on $\bar{\met}^{\mu\nu}$ must be explicitly revealed; for example, terms
which contain the trace $\tilde{h}$ must be rewritten as
$\bar\met_{\mu\nu}\tilde{h}^{\mu\nu}$. Indices should appear in their
``natural'' positions (see Sec.~\ref{sec:GaugeInAction}), 
for indices raised and lowered with the metric
have implicit dependence on it. Furthermore, since we are approaching
gravity from the metric formulation, rather than the Palatini
formulation, there will be contributions from the variation of
$\bcd_\mu$ and curvature tensors.

Consider a term in the effective action such as
\begin{equation}
S^{\mbox{\tiny{ex.1}}} = 
\int \rmd^4x \sqrt{-\bar\met}\  T^{\gamma\cdots}{}_{\delta\cdots}
\ \bcd_{\mu}\ S^{\alpha\cdots}{}_{\beta\cdots} \ ,
\end{equation}
where $\met$ is the determinant of the metric and $T^{\gamma\cdots}{}_{\delta\cdots},
S^{\alpha\cdots}{}_{\beta\cdots}$ are some tensor expressions, with all indices contracted in
some fashion, as the action must be a scalar. When such a term is
varied with $\bar\met^{\mu\nu} \to \bar\met^{\mu\nu} +
\delta\bar\met^{\mu\nu}$, besides the obvious contributions from
$\delta\sqrt{-\bar\met}$, and explicit dependence of $T^{\gamma\cdots}{}_{\delta\cdots},
S^{\alpha\cdots}{}_{\beta\cdots}$ on $\bar\met^{\mu \nu}$,
there are also contributions from varying the Christoffel
connection $\Gamma_{\mu \beta}^{\lambda}$ in $\bcd_\mu$. The general expression is
\begin{multline}
  \delta(\bcd_\mu S^{\alpha_1\cdots\alpha_n}{}_{\beta_1\cdots\beta_m}) = 
  \bcd_\mu(\delta S^{\alpha_1\cdots\alpha_n}{}_{\beta_1\cdots\beta_m}) + \\
  +\sum_{i=1}^n \delta\bar\Gamma^{\alpha_i}_{\mu\lambda}
  S^{\cdots\lambda\cdots}{}_{\beta_1\cdots\beta_m}
  -\sum_{j=1}^m \delta\bar\Gamma^\lambda_{\mu\beta_j}
  S^{\alpha_1\cdots\alpha_n}{}_{\cdots\lambda\cdots} \,,
\end{multline}
where $\cdots\lambda\cdots$ in the $i^{\rm th}$ term of a sum means
replacing $\alpha_i$ or $\beta_i$ in the index list with $\lambda$;
and where
\begin{equation}
  \delta\bar\Gamma^\sigma_{\mu\nu} = -\frac{1}{2}\left[
    \bar\met_{\lambda\mu} \bcd_\nu\delta\bar\met^{\lambda\sigma} + 
    \bar\met_{\lambda\nu} \bcd_\mu\delta\bar\met^{\lambda\sigma} -
    \bar\met_{\mu\alpha}\bar\met_{\nu\beta}\bcd^\sigma\delta\bar\met^{\alpha\beta}
  \right].
\end{equation}

Curvature tensors also depend on derivatives of the connection, so one naturally
expects terms of the form $\bcd_\rho\bcd_\sigma\delta\bar\met^{\mu \nu}$ from the variation
of curvature quantities, i.e.~the Riemann tensor $R^{\mu}{}_{\nu \alpha \beta}$, 
Ricci tensor $R_{\mu \nu}$, or Ricci scalar $R$. For example,
one can show that
\begin{equation}
\delta\bar R^\mu{}_{\nu\alpha\beta} =
2\bcd_{[\alpha}\delta\bar\Gamma^\mu_{\beta]\nu} \ ,
\end{equation}
where the contribution from $\Gamma \wedge \Gamma$ cancels~\cite{Carroll}.
Upon integration by parts any scalar in the action that contains curvature tensors, one can 
convert a term containing $\bcd_\rho\bcd_\sigma\delta\bar\met^{\mu\nu}$ into 
\begin{equation}
\label{eq:divgform}
\delta S^{\mbox{\tiny{ex.2}}} = 
\int\rmd^4x\  \sqrt{-\bar\met} \ P^\sigma{}_{\mu\nu} \bcd_\sigma
\delta\bar\met^{\mu\nu}\,.
\end{equation}
In fact, many terms in the variation of the action can be written in the form of Eq.~\eqref{eq:divgform}. 

The contribution of Eq.~\eqref{eq:divgform} to the effective stress-energy tensor is
found by integrating by parts and then averaging, according to
Eq.~\eqref{eq:Teffgeneral}. Upon averaging, however, one finds that such terms do
not contribute to the effective stress-energy tensor because
\begin{equation}
T^{\eff,\mbox{\tiny{ex.2}}}_{\mu\nu} =2\<\< \bcd_\sigma P^\sigma{}_{\mu\nu} \>\>\ ,
\end{equation}
vanishes according to property (2) in Sec.~\ref{sec:averageProps}. 

The above arguments and results imply that the variations of curvature tensors 
and connection coefficients with respect to $\bar\met_{\mu \nu}$ do not contribute to the 
effective stress-energy tensor. Only metric tensors which are raising, lowering, and
contracting indices in the action contribute to this tensor. We can thus concentrate
on these, when computing $T_{\mu \nu}^{\eff}$.

\subsection{Contributions at asymptotic infinity}
\label{sec:atscriplus}
When calculating the radiation-reaction force to leading order in the metric perturbation, 
it is crucial to account for all the energy-momentum loss in the system. The first contribution is
straightforward: energy-momentum is radiated outward, toward future, null infinity, $\scri^{+}$.
Since the stress-energy tensor is covariantly conserved, the energy-momentum radiated 
to $\scri^{+}$ can be calculated by performing a surface integral over a two-sphere at future, null
infinity\footnote{We will not consider spacetimes which are not asymptotically flat, e.g.~de~Sitter
space; the calculations are more involved in such spacetimes.}.

However, not all energy-momentum loss escapes to infinity, as
energy can also be lost due to the presence of trapped surfaces in 
the interior of the spacetime. Trapped surfaces can effectively absorb 
GW energy-momentum, which must also be accounted for, e.g.~in the
calculation of EMRI orbits around supermassive BHs~\cite{Yunes:2009ef,Yunes:2010zj}.
Calculations of such energy-momentum loss at the BH horizon are dramatically more 
complicated than those at $\scri^+$ and we do not consider them here.

What is the relative importance of energy-momentum lost to $\scri^{+}$
and that lost into trapped surfaces? To answer this question, we can concentrate
on the magnitude of the leading-order energy flux, as the argument trivially extends to 
momentum. The post-Newtonian (PN) approximation~\cite{Blanchet:2002av}, 
which assumes weak-gravitational fields and slow velocities, predicts that the  
energy flux carried out to $\scri^{+}$ is proportional to $v^{10}$ to leading-order in $v$, 
where $v$ is the orbital velocity of a binary system in a quasi-circular orbit 
(see e.g.~\cite{Blanchet:2002av}). On the other hand, a combination of the PN approximation 
and BH perturbation theory predicts that, to leading-order in $v$, the energy flux carried into 
trapped surfaces is proportional to $v^{15}$ for spinning BHs and $v^{18}$ for non-spinning 
BHs~\cite{Mino:1997p287}. BH GW flux absorption is then clearly smaller than the GW flux 
carried out to $\scri^{+}$ if $v <1$, which is true for EMRIs for
which the PN approximation holds.

Intuitively, this hierarchy in the magnitude of energy-momentum flux lost 
by BH binaries can be understood by considering the BH as a
geometric absorber in the radiation field. Radiation which is longer
in wavelength than the size of the BH is very weakly absorbed. Only at the end of an 
inspiral will the orbital frequency be high enough that GWs will be significantly absorbed 
by the horizon. Notice that this argument is independent of the particular theory considered, 
only relying on the existence of trapped surfaces. This result does not imply that BH absorption
should be neglected in EMRI modeling, but just that it is a smaller effect than the flux carried out to 
infinity~\cite{Yunes:2009ef,Yunes:2010zj}.

In the remainder of the paper, we will only address energy-momentum radiated to $\scri^{+}$
and relegate any analysis of radiation lost into trapped surface to future work.
The only terms which can contribute to an energy-momentum flux integral on a
2-sphere at $\scri^+$ are those which decay as $r^{-2}$, since the
area element of the sphere grows as $r^2$. No terms may decay more
slowly than $r^{-2}$, as the flux must be finite, i.e.~the effective stress-energy cannot
scale as $r^{-1}$, as a constant or with positive powers of radius. 
Similarly, any terms decaying faster 
than $r^{-2}$ do not contribute, as they would vanish at $\scri^{+}$. Of course, to determine
which terms contribute and which do not, one must know the leading asymptotic forms of all quantities in the effective stress-energy tensor.

In GR, as we shall see in \autoref{sec:GRTab}, the only fields
appearing in the effective stress-energy tensor are the background
metric $\bar\met_{\mu \nu}$ and derivatives of the metric perturbation
$\tilde{h}_{\mu \nu}$. As one approaches $\scri^{+}$, $\bar\met_{\mu
  \nu} \sim \eta_{\mu \nu}$ in Cartesian coordinates, while
$|\tilde{h}_{\mu \nu}| \sim r^{-1} \sim |\bcd_{\rho}\tilde{h}_{\mu
  \nu}|$. Curvature tensors scale as $|\overline{R}_{\mu \nu \delta
  \sigma}|\sim r^{-3}$, since they quantify tidal forces.  For a
theory that is a deformation away from GR, and far away from regions
of strong curvature, these asymptotic forms cannot change.

Consider now terms in the general effective action at order $\scO(\epsilon^2)$
that contain background curvature tensors. Due to their ordering, they would
contribute to the effective stress-energy. One such term is
\begin{equation}
\label{eq:curvatureExample}
S^{\mbox{\tiny{ex.3}}} =
\epsilon^2 \int \rmd^4x \sqrt{-\bar\met}\ \bcd_\rho \tilde{\varphi}_1^\sigma
\ \bcd_\alpha \tilde{\varphi}_2^\beta\ \overline{R}^\rho{}_{\beta
  \sigma\kappa} \ \bar\met^{\alpha\kappa}\ ,
\end{equation}
where $\tilde{\varphi}_1,\tilde{\varphi}_2$ are the first-order
perturbations to two fields in the theory (e.g.~the metric
perturbation $\tilde{h}_{\mu\nu}$ and the CS scalar perturbation
$\tilde\vartheta$ that we introduce in \autoref{sec:CSTab}).
Since there is no contribution to the effective stress-energy tensor from the variation
of curvature quantities (see \autoref{sec:varR}), the only contributions to the effective 
stress-energy comes from 
\begin{multline}
T^{\eff,{\mbox{\tiny{ex.3}}}}_{\mu\nu} = 
-2\epsilon^2 \left\< \left\< \left( -\frac{1}{2}
  \bar\met_{\mu\nu}\bar\met^{\alpha\kappa} +
  \delta^\alpha{}_{(\mu}\delta^\kappa{}_{\nu)}\right) \cdot
\right.\right.\\
\cdot \underbrace{\bcd_\rho \tilde{\varphi}_1^\sigma}_{r^{-1}}
\ \underbrace{\bcd_\alpha \tilde{\varphi}_2^\beta}_{r^{-1}}\
\underbrace{\overline{R}^\rho{}_{\beta\sigma\kappa}}_{r^{-3}}
\Big\> \Big\>\,,
\label{toy-term}
\end{multline}
which has the same functional form as the integral.
Note that the curvature tensor always remains when varying with respect
to $\bar\met^{\mu \nu}$.

Combining this result with the asymptotic arguments above, such terms can be
ignored as one approaches $\scri^+$. Each of the first-order fields possess a
radiative part that scales as $r^{-1}$. The square of the first-order fields would then
satisfy the $r^{-2}$ scaling requirement for the flux integral. The curvature tensor, however,
scales as $r^{-3}$, which implies that the term in Eq.~\eqref{toy-term} vanishes at $\scri^{+}$.

We then conclude that terms in the action that contain background 
curvature quantities at $\scO(\epsilon^2)$ may be ignored in calculating the effective 
stress-energy tensor at $\scri^+$. 
As an immediate corollary to this simplification, we may also freely commute background 
covariant derivatives if we are interested in the stress-energy tensor at
infinity only, since the commutator is proportional to background curvature
tensors.

\subsection{Imposing gauge in the effective action}
\label{sec:GaugeInAction}
We will choose as our dynamical field not $\tilde{h}_{\mu\nu}$ but
rather $\barh^{\mu\nu}$, where the underline stands for the
trace-reverse operation, and we take the ``natural'' position of the
indices to be contravariant. The resulting stress-energy tensor is
equal to the one calculated using $\tilde{h}_{\mu\nu}$ \emph{after}
evaluating both of them on-shell, i.e.~imposing the equations of
motion. 

We also impose a gauge condition to simplify future
expressions: the Lorenz gauge condition,
\begin{equation}
\label{eq:LorenzGauge}
\bcd_\mu \barh^{\mu\nu} = 0\,.
\end{equation}
Typically one may not impose a gauge condition at the level of the
action. However, in our case,
the gauge condition in Eq.~\eqref{eq:LorenzGauge} has the important
property of having all of the indices in their natural positions:
the contraction of the indices does not involve the
metric. 

Consider a term in the effective action that contains this
divergence,
\begin{equation}
S^{\mbox{\tiny{ex.4}}} = \epsilon^2\int \rmd^4x \sqrt{-\bar\met}\ T_\beta
\ \bcd_\alpha\barh^{\alpha\beta}\, ,
\end{equation}
with $T_\beta$ some tensor expression at first order in $\epsilon$.
The $\alpha$ index that is contracted above does not require
the metric for such contraction. Therefore, $\bcd_\alpha \barh^{\alpha\beta}$ 
always remains upon variation,
\begin{equation}
T^{\eff,\mbox{\tiny{ex.4}}}_{\mu\nu} = -2\epsilon^2 
\left\<\left\< \left( -\frac{1}{2}\bar\met_{\mu\nu}T_\beta + \frac{\delta
  T_\beta}{\delta \bar\met^{\mu\nu}} \right)
\bcd_\alpha \barh^{\alpha\beta} \right\>\right\>\ .
\end{equation}

If we delayed imposing the Lorenz gauge condition until after the
calculation of the effective stress-energy tensor, we would find the
same effective tensor as if we had imposed the gauge condition at the
level of the action. Having said that, one should not impose the gauge condition when
varying with respect to $\barh^{\mu\nu}$ as clearly $\bcd_\alpha\barh^{\alpha\beta}$ must
also be varied. 

\section{Effective Stress-Energy in GR}
\label{sec:GRTab}

Let us now demonstrate the principles described in the previous section by deriving the standard
Isaacson stress-energy tensor in GR. Consider the Einstein-Hilbert action,
\begin{equation}
\label{eq:SEH}
S_\GR = \kappa \int \rmd^4x \sqrt{-\met} \; R,
\end{equation}
where $\kappa = (16\pi G)^{-1}$. Now
perturb to second order to form the effective action,
\begin{subequations}
\label{eq:SGReff}
\begin{equation}
S_\GR^\eff = S_\GR^{\eff(0)} + S_\GR^{\eff(2)},
\end{equation}
\begin{eqnarray}
S_\GR^{\eff(0)} &=& \kappa \int \rmd^4x \sqrt{-\bar\met} \  \bar{R}, \\
S_\GR^{\eff(2)} &=& \epsilon^2 \kappa \int\rmd^4x 
\ \scL_\GR^{\eff,1}+\scL_\GR^{\eff,2} \ ,
\end{eqnarray}
where
\begin{multline}
\label{eq:LeffGR1}
\scL_\GR^{\eff,1} = \frac{1}{8} \sqrt{-\bar\met} \left[ 4\bar{R}_{\alpha\beta} \left( 2
    \barh^\alpha{}_\mu \barh^{\beta\mu} -
    \barh^{\alpha\beta}\barh \right)\right. \\
  \left.+ \bar{R} \left(\barh^2 - 2
    \barh_{\alpha\beta} \barh^{\alpha\beta} \right) \right]\ ,
\end{multline}
and
\begin{multline}
\label{eq:LeffGR2}
\scL_\GR^{\eff,2} = \sqrt{-\bar\met}\left[-\barh^{\alpha\beta} \bcd_\alpha
\bcd_\mu \barh^\mu_{\phantom{\mu}\beta}- \frac{1}{8} (\bcd_\mu
\barh)(\bcd^\mu \barh)- \right.\\
(\bcd_\mu \barh^\mu_{\phantom{\mu}\alpha})(\bcd_\nu \barh^{\nu\alpha})
+ \frac{1}{2} (\bcd_\nu \barh)(\bcd_\mu \barh^{\mu\nu})
-\barh^{\alpha\beta}
\bcd_\mu\bcd_\alpha\barh^\mu_{\phantom{\mu}\beta} \\
+\frac{1}{2}\barh\bcd_\mu\bcd_\nu\barh^{\mu\nu}
+\barh^{\alpha\beta} \bar{\square}\barh_{\alpha\beta}
- \frac{1}{4} \barh\bar{\square}\barh \\
\left.
-\frac{1}{2} (\bcd_\mu\barh_{\nu\alpha})
(\bcd^\nu\barh^{\mu\alpha})
+\frac{3}{4}(\bcd_\mu\barh_{\alpha\beta}) (\bcd^\mu
\barh^{\alpha\beta}) \right] \,,
\end{multline}
\end{subequations}
and $(\bar{R}_{\mu \nu},\bar{R})$ refer to the background Ricci tensor and scalar respectively. 
The integrands have been written in terms of the trace-reversed metric
perturbation, $\barh^{\mu \nu}$. From \autoref{sec:atscriplus},
$\scL_\GR^{\eff,1}$ does not contribute at $\scri^+$ because it depends explicitly
on curvature quantities, so we ignore it. The variation and averaging
of $\scL_\GR^{\eff,2}$ produces the Isaacson stress-energy tensor.

By integrating by parts, all terms in Eq.~\eqref{eq:LeffGR2} can be
written as $(\bcd_\alpha\barh^{\rho\sigma})(\bcd_\beta\barh^{\kappa\lambda})$
(with indices contracted to form a scalar) rather than
$\barh^{\rho\sigma}\bcd_\alpha\bcd_\beta\barh^{\kappa\lambda}$ (again,
with indices contracted). $\scL_\GR^{\eff,2}$ is thus rewritten in the more
compact form
\begin{multline}
\label{eq:LMT}
\scL_\MT = \sqrt{-\bar\met}\left[\frac{1}{2} (\bcd_\mu
  \barh^{\nu\alpha}) (\bcd_\nu \barh^\mu{}_\alpha) - \right.\\
\left.  -\frac{1}{4} (\bcd_\mu \barh_{\alpha\beta}) (\bcd^\mu
  \barh^{\alpha\beta}) + \frac{1}{8} (\bcd_\mu \barh) (\bcd^\mu \barh)
\right]\ ,
\end{multline}
which is the expression that appears in MacCallum and Taub~\cite{1973CMaPh..30..153M}. With this simplified
expression at hand, we can promote $\barh^{\mu \nu}$ to an independent dynamical field 
in Eq.~\eqref{eq:SGReff} and vary it with respect to both $\bar\met^{\mu \nu}$ and
$\barh^{\mu \nu}$ to obtain the effective stress-energy tensor and the first-order equations
of motion respectively. 

Let us first derive the first-order equations of motion. 
Varying Eq.~\eqref{eq:SGReff} with respect to $\barh^{\mu \nu}$, we find
\begin{subequations}
  \label{eq:GRGWEOM}
  \begin{equation}
    \label{eq:GRGWEOMtens}
    \bar{\square} \barh_{\mu\nu} -2 \bcd^\alpha\bcd_{(\mu}
    \barh_{\nu)\alpha} -\frac{1}{2}\bar\met_{\mu\nu} \bar{\square} \barh =
    0\ ,
  \end{equation}
whose trace is
  \begin{equation}
  \label{eq:GRGWEOMtr}
    2\bcd_\alpha\bcd_\beta\barh^{\alpha\beta}+\bar{\square}\barh = 0\ .
  \end{equation}
We can now impose the Lorenz gauge on Eq.~\eqref{eq:GRGWEOMtr}, which then 
leads to $\bar{\square}\barh = 0$. If $\barh = 0$ is further imposed on an initial hypersurface while
maintaining Lorenz gauge, then the evolution equation preserves the
trace-free gauge~\cite{Misner:1973cw}. The combination of these two
gauge choices (Lorenz gauge plus trace-free) is the transverse-tracefree gauge, or TT gauge.

After commuting derivatives in Eq.~\eqref{eq:GRGWEOMtens} and imposing TT
gauge, the tensor equation of motion reads
\begin{equation}
\bar{\square}\barh_{\mu\nu} + 2\bar R_{\mu\alpha\nu\beta}
\barh^{\alpha\beta} = 0\,,
\end{equation}
\end{subequations}
where $\bar{R}_{\mu \alpha \nu \beta}$ is the background Riemann tensor.
At $\scri^+$, this equation reduces to $\bar{\square}\barh_{\mu\nu}=0$, 
which leads to the standard dispersion relation for GWs, traveling at the speed of light.

Let us now calculate the effective stress-energy tensor.
Note that the first term in $\scL_\MT$ may be integrated by parts and
covariant derivatives commuted to form the Lorenz gauge condition, so
the first term may be ignored. Varying the action with respect to
$\bar\met^{\mu \nu}$, we find
\begin{subequations}
\begin{equation}
\kappa G_{\mu\nu} = -\epsilon^2\kappa \left\<\left\< \frac{1}{\sqrt{-\bar\met}}
\frac{\delta}{\delta\bar\met^{\mu\nu}} \scL_\MT \right\>\right\>
\equiv \frac{1}{2} T_{\MT\mu\nu}^\eff \ ,
\end{equation}
where
\ba
\label{eq:TMT}
T_{\MT\mu\nu}^\eff &=& 
2\epsilon^2\kappa \left\< \left\< 
\frac{1}{4} \bcd_\mu \barh^{\alpha\beta} \bcd_\nu \barh_{\alpha\beta}
-\frac{1}{2} \bcd_\alpha \barh_{\beta\mu} \bcd^\alpha \barh_\nu{}^\beta 
\right.\right.\nonumber \\
&&\qquad{}- \frac{1}{8} \bcd_\mu \barh \bcd_\nu \barh
+\frac{1}{4} \bcd_\alpha \barh_{\mu\nu} \bcd^\alpha \barh 
\nonumber \\ 
&&\qquad\left.\left.{}+ \frac{1}{2} \bar\met_{\mu\nu} (-\bar\met)^{-1/2} \scL_\MT \right\>\right\> \,,
\ea
which we refer to as the MacCallum-Taub tensor. Terms that depend on the trace $\barh^\mu{}_\mu$ in this tensor can be eliminated in TT gauge.

Let us now evaluate the MacCallum-Taub tensor on
shell, by imposing the equations of motion [Eq.~\eqref{eq:GRGWEOM}]. When
short-wavelength averaging, derivatives that are contracted
together can be converted into the d'Alembertian via integration by parts; such
terms vanish at $\scri^+$. What results is the usual Isaacson stress-energy tensor,
\begin{equation}
T_{\GR\mu\nu}^\eff = \epsilon^2\frac{\kappa}{2}
\left\<\left\< \left(\bcd_\mu \barh^{\alpha\beta}\right)
\left(\bcd_\nu \barh_{\alpha\beta} \right) \right\>\right\>\,.
\end{equation}
\end{subequations}
Notice that this expression is only valid at $\scri^+$ and in TT gauge. As
mentioned earlier, this tensor cannot be used to model energy-momentum loss
through trapped surfaces, since then curvature quantities
cannot be ignored.

\section{Chern-Simons Gravity}
\label{sec:CS}

CS gravity is a modified theory introduced first by Jackiw and Pi~\cite{jackiw:2003:cmo} (for a recent review see~\cite{Alexander:2009tp}). The dynamical version of this theory modifies the Einstein-Hilbert action through the addition of the following terms:
\ba
\label{full-action}
S &=& S^{}_{\rm EH} + S^{}_{\rm CS} +  S^{}_{\vartheta} + S^{}_{\rm mat},
\ea
where
\begin{eqnarray}
\label{actions}
S^{}_\EH &=& \kappa \int d^4x  \; \sqrt{-\met}  \; R\,, 
\nonumber \\
S^{}_\CS &=& \frac{\alpha}{4} \int d^4x \; \sqrt{-\met} \quad
\vartheta \; \pont\,,
\nonumber \\
S^{}_{\vartheta} &=& - \frac{\beta}{2} \int d^{4}x \; \sqrt{-\met} \;  \met^{\mu \nu}
\left(\nabla_{\mu} \vartheta\right) \left(\nabla_{\nu} \vartheta\right) \,,
\nonumber \\
S^{}_{\textrm{mat}} &=& \int d^{4}x \; \sqrt{-\met} \; {\cal{L}}_{\textrm{mat}}\,.
\end{eqnarray}
The quantity $\kappa = (16 \pi G)^{-1}$ is the gravitational constant, while $\alpha$ and $\beta$ are coupling constants that control the strength of the CS coupling to the gravitational sector and its kinetic energy respectively. In the non-dynamical version of the theory, $\beta = 0$ and there are no dynamics for the scalar field, which is promoted to a prior-geometric quantity.

The quantity $\vartheta$ is the CS field, which couples to the
gravitational sector via the parity-violating Pontryagin density,
$\pont$, which is given by
\be
\pont  := R^{}_{\alpha \beta \gamma \delta}
{\,^\ast\!}R^{\alpha \beta \gamma \delta} =
\frac{1}{2} \varepsilon^{\alpha \beta \mu \nu} 
R^{}_{\alpha \beta \gamma \delta}
R^{\gamma \delta}{}^{}_{\mu \nu}\,,
\label{CSmodification}
\ee
where the asterisk denotes the dual tensor, which we construct using the antisymmetric Levi-Civita tensor $\varepsilon^{\alpha\beta\mu\nu}$. This scalar is a topological invariant, as it can
be written as the divergence of a current
\begin{equation}
\label{eq:pontdiv}
\pont = 4 \cd_\mu \left[ \varepsilon^{\mu\alpha\beta\gamma}
  \Gamma^\sigma_{\alpha\tau}
\left( \frac{1}{2} \pd_\beta \Gamma^\tau_{\gamma\sigma} +
  \frac{1}{3} \Gamma^\tau_{\beta\eta}\Gamma^\eta_{\gamma\sigma}
\right) \right] \,.
\end{equation}

Equation~\eqref{full-action} contains several terms that we describe below: the first one is the Einstein-Hilbert action; the second one is the CS coupling to the gravitational sector;  the third one is the CS kinetic term; and the fourth one stands for additional matter degrees of freedom.  
The CS kinetic term is precisely the one that distinguishes the non-dynamical and the dynamical theory. In the former, the scalar field is {\emph{a priori}} prescribed, while in the dynamical theory, the scalar field satisfies an evolution equation. 
 
The field equations of this theory are obtained by varying the action with respect to all degrees of freedom:
\begin{subequations}
\label{eq:EOMs}
\ba
G^{}_{\mu \nu} + \frac{\alpha}{\kappa} C^{}_{\mu \nu}  &=& \frac{1}{2 \kappa} \left(T_{\mu \nu}^{\MAT} 
+ T_{\mu \nu}^{(\vartheta)}\right)\,,  \label{EEs} \\
\beta \square \vartheta &=& - \frac{\alpha}{4} \pont\,,
\label{eq:EOMtheta0}
\ea
\end{subequations}
where $T_{\mu \nu}^{\MAT}$ is the matter stress-energy tensor and $T_{\mu \nu}^{(\vartheta)}$ is the CS scalar stress-energy: 
\be
T_{\mu \nu}^{(\vartheta)} = \beta \left[ (\nabla_{\mu} \vartheta) (\nabla_{\nu}\vartheta) - \frac{1}{2} 
\met_{\mu \nu}(\nabla^{\sigma} \vartheta) (\nabla_{\sigma} \vartheta) \right]\,.
\label{eq:vartheta-Tab}
\ee
The C-tensor $C^{\mu \nu}$ is given by
\be
\label{Ctensor}
C^{\alpha \beta} = \left(\nabla^{}_{\sigma} \vartheta\right) \varepsilon^{\sigma \delta \nu(\alpha}
\nabla^{}_{\nu}R^{\beta)}{}^{}_{\delta} +  \left(\nabla^{}_{\sigma}\nabla^{}_{\delta}\vartheta \right) 
{\,^\ast\!}R^{\delta (\alpha \beta)\sigma}\,.
\ee

Many solutions to these field equations have been found. In their pioneering work, Jackiw and Pi showed that the Scwharzschild metric is also a solution in CS gravity~\cite{jackiw:2003:cmo}. Later on, a detailed analysis showed that all spherically symmetric spaces, such as the Friedman-Robertson-Walker metric, are also solutions~\cite{Grumiller:2007rv}. Axially-symmetric spaces, however, are not necessarily solutions, because the Pontryagin density does not vanish in this case, sourcing a non-trivial scalar field. Specifically, this implies the Kerr metric is not a solution.

A slowly-rotating solution, however, does exist in dynamical CS gravity. Yunes and Pretorius~\cite{2009PhRvD..79h4043Y} found that when the field equations are expanded in the Kerr parameter $a/M \ll1$ and in the small-coupling parameter $\zeta \equiv \xi/M^{4} = \alpha^{2}/(\beta \kappa M^{4}) \ll 1$, then the CS field equations have the solution
\ba
\label{eq:slowRot}
d\bar{s}^{2} &=& ds^{2}_{\rm Kerr} + \frac{5}{8} \zeta \frac{M a}{r^{4}} \left(1 + \frac{12}{7} \frac{M}{r}  + \frac{27}{10} \frac{M^{2}}{r^{2}} \right) \sin^{2}{\theta} dt d\phi\,, 
\nonumber \\
\bar{\vartheta} &=& \frac{5}{8}\frac{\alpha}{\beta}\frac{a}{M} \frac{\cos\theta}{r^2}
\left( 1 + \frac{2M}{r} + \frac{18M^2}{5r^2} \right)\,,
\ea
to second order in $a/M$ and to first order in $\zeta$, assuming no matter sources.  These
equations employ Boyer-Lindquist coordinates $(t,r,\theta,\phi)$ and
$ds^{2}_{\rm Kerr}$ is the Kerr line element. The solution for
$\vartheta$ may also include an arbitrary additive constant, but this
constant is unimportant, since only derivatives of $\vartheta$ enter the CS field equations. Recently, the same solution has been found to linear order in $a/M$ in the Einstein-Cartan formulation of the non-dynamical theory~\cite{Cambiaso:2010un}.

The divergence of the field equations reduce to $\nabla^{\mu} T_{\mu \nu}^{\MAT} = 0$. This is because the divergence of the Einstein tensor vanishes by the Bianchi identities. Meanwhile, the divergence of the C-tensor exactly cancels the divergence of the CS scalar field stress-energy tensor, upon imposition of the equations of motion [Eq.~\eqref{eq:EOMtheta0}]. Therefore, test-particle motion in dynamical CS gravity is exactly geodesic\footnote{This statement is true only in the absence of spins, since otherwise the CS effective worldline action would contain new self-interaction terms.}. This result automatically implies the weak-equivalence principle is satisfied.

The gravitational perturbation only possesses two independent, propagating degrees of freedom or polarizations. Jackiw and Pi showed that this was the case in the non-dynamical theory~\cite{jackiw:2003:cmo}, while Sopuerta and Yunes did the same in the dynamical version~\cite{Sopuerta:2009iy}. One can also show easily that a transverse and approximately traceless gauge exists in dynamical CS gravity. The trace of the field equations take the interesting form
\be
- R = \frac{1}{2 \kappa} \left(T^{\MAT}  + T^{(\vartheta) }\right)\,, 
\label{eq:pre-trace}
\ee
where $R$ is the Ricci scalar and $T$ is the trace of the stress-energy tensor. Notice that the trace of the C-tensor vanishes identically. 

In vacuum ($T^\MAT_{\mu\nu} = 0$) and when expanding to linear order about a Minkowski background, Eq.~\eqref{eq:pre-trace} reduces to
\be
\bar{\square} {\tilde{h}} = \frac{\beta}{2 \kappa} (\bcd^{\sigma} \bar\vartheta) (\bcd_{\sigma} \bar\vartheta) \,.
\label{eq:trace-free}
\ee
where $\tilde{h}\equiv \eta^{\mu \nu} h_{\mu \nu}$ is the trace of the
metric perturbation, $\bar{\square}$ is the d'Alembertian operator
with respect to the background metric and $\bar{\vartheta}$ is the
background scalar field. Since the latter must satisfy the evolution
equation [Eq.~\eqref{eq:EOMtheta0}], we immediately see that
$\vartheta \propto \alpha/\beta$. This means that the right-hand side
of Eq.~\eqref{eq:trace-free} is proportional to $\zeta$. To zeroth
order in the small-coupling approximation, $\tilde{h}$ then satisfies
a free wave equation and can thus be treated as vanishing. Deviations
from the trace-free condition can only arise at ${\cal{O}}(\zeta)$ and
they are suppressed by factors of the curvature tensor, as
$\bar\vartheta$ must satisfy Eq.~\eqref{eq:EOMtheta0}. Approaching
$\scri^+$, the right hand side of Eq.~\eqref{eq:trace-free}
vanishes. This allows one to impose TT gauge at future null infinity.

\section{Effective Stress-Energy in CS Gravity}
\label{sec:CSTab}

The perturbed Lagrangian for the Einstein-Hilbert action has already been calculated, so here
we need only consider the contribution from $S_\CS$. At 
$\scO(\epsilon^2)$, there are a large number of terms generated (we
used the package \emph{xPert}~\cite{2007CoPhC.177..640M,2008CoPhC.179..586M,2008CoPhC.179..597M,2009GReGr..41.2415B,Xact} to calculate the
perturbations). Many of these terms are irrelevant when considering
their contribution at $\scri^+$. 

Let us classify the types of terms that arise in $S_{\CS}$. At $\scO(\epsilon^2)$, these are of two types: 
\begin{enumerate}
\item the second-order part of one field, or 
\item the product of first-order parts of two fields.
\end{enumerate}
As mentioned in \autoref{sec:averageProps}, terms containing the second-order part of one
field are linear in a short-wavelength quantity, which vanishes under
averaging. Thus we only need to consider the
latter case. There are five fields in $S_\CS$ ($\sqrt{-\met},\ 
\varepsilon,\ \vartheta,\ R$, and $\starR$), so one would at first
think that there are $\binom{5}{2}=10$ types of terms arising; however,
from the definition of the Levi-Civita tensor, 
we have
\begin{equation}
\label{eq:LeviCivitaDef}
\sqrt{-\met} \; \varepsilon^{\alpha\beta\mu\nu} = \sign(\met)\left[\alpha\beta\mu\nu\right],
\end{equation}
where $\left[\alpha\beta\mu\nu\right]$ is the Levi-Civita
\emph{symbol}, which is not a spacetime field. The combination
$\sqrt{-\met} \; \varepsilon^{\alpha\beta\mu\nu}$ therefore has no perturbation, and
there are only three spacetime fields which contribute. We are left with
only $\binom{3}{2}=3$ possibilities for the types of terms
that could appear, corresponding to two perturbed fields and one unperturbed
one among the set $(\vartheta,R_{\alpha \beta \sigma \delta},R_{\alpha \beta \sigma \delta})$. 
Two of these possibilities are actually the same by exchanging the 
two copies of $R_{\alpha \beta \sigma \delta}$.

Therefore, we are left with only the following two types of terms in the CS Lagrangian density:
\begin{subequations}
\label{eq:pertCSLterms}
\begin{eqnarray}
\scL_{\tilde{\vartheta} \tilde{R}}&\sim&\sqrt{-\bar\met} \; \bar\varepsilon^{\alpha\beta\mu\nu}
\ \tildetheta \ \tilde{R}^\gamma{}_{\delta\alpha\beta}
\bar{R}^\delta{}_{\gamma\mu\nu} \label{eq:pertCSLterms1}\\
\scL_{\tilde{R}\tilde{R}}&\sim&\sqrt{-\bar\met} \; \bar\varepsilon^{\alpha\beta\mu\nu}
\ \bar\vartheta \ \tilde{R}^\gamma{}_{\delta\alpha\beta}
\tilde{R}^\delta{}_{\gamma\mu\nu} \label{eq:pertCSLterms2}
\end{eqnarray}
\end{subequations}
where $\tilde{R}^\gamma{}_{\delta\alpha\beta}$ is the first-order perturbation to the Riemann
tensor, in terms of $\barh^{\mu \nu}$. Variation of these terms with respect to the background
metric yields the CS contributions to the effective stress-energy tensor, while variation with
respect to the metric perturbation yields CS corrections to the first
order equations of motion.

\subsection{Variation with respect to the Perturbation}

Just as in GR, the final expression for the stress-energy tensor must be put on-shell by imposing
the equations of motion. The first-order equations of motion of
dynamical CS gravity, in vacuum and at $\scri^+$, are
\begin{multline}
\label{eq:EOMCS1atscri}
\bar\square\barh_{\mu\nu} = -\frac{1}{\kappa}\tilde{T}^{(\vartheta)}_{\mu\nu} +
\frac{\alpha}{\kappa} \left[ 
\bcd_\alpha\bar\vartheta \ \bcd_\beta \bar\square \barh_{\gamma(\mu}
\ \bar\varepsilon^{\alpha\beta\gamma}{}_{\nu)}
\right. \\
\left. +
\bcd_\alpha\bcd_\beta\bar\vartheta \
\bar\varepsilon^\alpha{}_{\gamma\delta(\mu} \ \bcd^\delta
\left(  \bcd_{\nu)}\barh^{\beta\gamma} -
  \bcd^\beta\barh_{\nu)}{}^\gamma \right) 
\right]\,.
\end{multline}

Imposing these equations of motion is easier when taking
advantage of the weak-coupling limit, $\zeta_{\GW}\ll 1$, where $\zeta_{\GW} \equiv
\alpha\bcd\vartheta/(\kappa \lambda_{\GW})$ quantifies the size of the deformation
away from GR. Let us then expand the metric perturbation in a Taylor series
\begin{equation}
\label{eq:gen-alt-h-exp-old}
\barh_{\mu\nu} = \sum_{n=0}^\infty (\zeta_{\GW})^n \barh^{(n)}_{\mu\nu}\,.
\end{equation}
To zeroth-order, it is clear that Eq.~\eqref{eq:EOMCS1atscri} reduces to 
$\bar\square \barh_{\mu \nu}^{(0)} = 0$, which is the standard GR equation of motion.
To next order, the the leading-order piece of the right-hand side vanishes and one is
then left with
\begin{eqnarray}
\label{eq:EOMCS1atscriagain}
\bar\square\barh_{\mu\nu}^{(1)} &=& 
\frac{\alpha}{\kappa}\ 
\bcd_\alpha\bcd_\beta\bar\vartheta \
\bar\varepsilon^\alpha{}_{\gamma\delta(\mu} \ \bcd^\delta
\left(  \bcd_{\nu)}\barh^{\beta\gamma}_{(0)}
 - \bcd^\beta\barh_{\nu)}{}^\gamma_{(0)} \right)\nn\\
&&-\frac{1}{\kappa}\tilde{T}^{(\vartheta)}_{\mu\nu}\,.
\end{eqnarray}
In the remainder of this section, we drop the superscripts that indicate $\zeta_{\GW}$-ordering.

\subsection{Variation with respect to the Background}

Let us first discuss terms of type $\scL_{\tildetheta \tilde{R}}$
under variation with respect to $\bar\met^{\mu\nu}$.  From \autoref{sec:varR}, 
only total derivative terms arise from
$\delta\bar{R}^\gamma{}_{\delta\alpha\beta}$, 
and these vanish upon averaging. The remaining terms contain
$\bar{R}^\gamma{}_{\delta\alpha\beta}$, which must vanish at 
$\scri^+$. Thus, as mentioned before, terms in the effective
action which contain curvature tensors do not contribute to the
effective stress-energy tensor at $\scri^+$.

We are then only left with $\scL_{\tilde{R}\tilde{R}}$.
Writing these in terms of $\barh^{\mu \nu}$, the effective
action reads
\begin{equation}
\label{eq:CSeff2}
S_\CS^{\eff(2)} = \epsilon^2\frac{\alpha}{4} \int \rmd^4x \ 
\scL_\CS^{\eff,1} + \scL_\CS^{\eff,2} \ ,
\end{equation}
where
\begin{subequations}
\label{eq:CSeffS}
\begin{eqnarray}
\label{eq:CSLeff1}
\scL_\CS^{\eff,1} &=& +\sqrt{-\bar\met} \;
\bar\varepsilon^{\alpha\beta\gamma\delta} \ \bar\vartheta
\ \bcd^\rho\bcd_\beta\barh _\alpha{}^\sigma \ 
\bcd_\delta\bcd_\rho\barh_{\sigma\gamma}\ ,\\
\label{eq:CSLeff2}
\scL_\CS^{\eff,2} &=& - \sqrt{-\bar\met} \;
\bar\varepsilon^{\alpha\beta\gamma\delta} \ \bar\vartheta
\ \bcd_\beta\bcd_\rho\barh_\alpha{}^\sigma \ 
\bcd_\sigma\bcd_\delta \barh^\rho{}_\gamma \ .
\end{eqnarray}
\end{subequations}
Naively, one might think that these expressions lead to an effective
stress-energy tensor at $\scO(\lambda_\GW^{-4})$. This is premature, however, 
as there can be a cancellation of $\lambda_\GW^{-4}$-terms that lead to a less steep
wavelength dependence. One should try to move as many
derivatives away from the perturbed quantities as possible before
proceeding. In fact, we know that this must be possible
from~\cite{jackiw:2003:cmo}: the Pontryagin density can be
written as the divergence of a 4-current, so at least one derivative
can be moved off of $\barh^{\mu \nu}$. This automatically implies that there cannot
be $\lambda_\GW^{-4}$ terms in the effective stress-energy tensor, as shown explicitly by
Yunes and Sopuerta~\cite{Sopuerta:2009iy}.

Let us transform $\scL_\CS^{\eff,1}$ in the following way.
The Levi-Civita tensor is contracted onto two
derivative operators ($\bcd_\beta$ and $\bcd_\delta$). One may
integrate by parts to move one of these derivative operators onto the remaining terms
in Eq.~\eqref{eq:CSLeff1}.
This generates two types of terms: one with three derivatives 
acting on the metric perturbation and one with one derivative on the CS scalar (the term
acting on the Levi-Civita tensor or the determinant of the metric vanishes by metric
compatibility). Let us focus on the former first.  
Because of the contraction onto the Levi-Civita tensor, 
only the antisymmetric part of the second derivative operator would contribute. 
Such a combination is nothing but the commutator of covariant derivatives, which can 
be written as the Riemann tensor, and thus vanishes at $\scri^{+}$. 
The remaining term with a covariant derivative of the CS scalar does not generically vanish. 
Dropping terms proportional to the Riemann tensor, $\scL_\CS^{\eff,1}$ becomes
\begin{subequations}
\label{eq:CSLintegrated}
\begin{equation}
\label{eq:CSLeff1integrated}
\scL_\CS^{\eff,1} = 
\sqrt{-\bar\met}
\bar\varepsilon^{\alpha\beta\gamma\delta} \ \bcd_\alpha\bar\vartheta \ 
\bcd^\rho\barh_\beta{}^\sigma
\bcd_\delta \bcd_{\rho} \barh_{\sigma\gamma} 
\ .
\end{equation}

Equation \eqref{eq:CSLeff2} can be analyzed with the property 
discussed in \autoref{sec:GaugeInAction}: Lorenz gauge may be imposed at the level
of the action for the purposes of calculating the effective
stress-energy tensor. This means that if one integrates by parts, moving $\bcd_\sigma$
and $\bcd_\rho$ onto remaining terms, the only term that survives is proportional to 
$\bar\vartheta$, as the divergence of $\barh^{\mu \nu}$ vanishes 
(after commuting derivatives, dropping Riemann terms and imposing Lorenz gauge). 
Thus $\scL_\CS^{\eff,2}$ becomes
\begin{equation}
\label{eq:CSLeff2integrated}
\scL_\CS^{\eff,2} = \sqrt{-\bar\met}
\bar\varepsilon^{\alpha\beta\gamma\delta} \ 
\bcd_\rho \bcd_\sigma \bar\vartheta \ 
\bcd_\alpha \barh_\beta{}^\sigma
\bcd_\gamma \barh_\delta{}^\rho
\ .
\end{equation}
\end{subequations}

With these simplified Lagrangian densities at hand, we can now compute the total effective stress-energy tensor for GWs in CS gravity:
\begin{equation}
\label{eq:CSeffTtotal}
T_{\CS\mu\nu}^\eff = T_{\MT\mu\nu}^\eff + T_{\CS\mu\nu}^{\eff,1}
+T_{\CS\mu\nu}^{\eff,2}\ ,
\end{equation}
where $T_{\CS\mu\nu}^{\eff,1}$ and $T_{\CS\mu\nu}^{\eff,2}$ are due to the variation of
$\scL_\CS^{\eff,1}$  and $\scL_\CS^{\eff,2}$ respectively. These expressions are
\begin{subequations}
\label{eq:CSeffT}
\begin{eqnarray}
\label{eq:CSeffT1}
T_{\CS\mu\nu}^{\eff,1} &=& -\epsilon^2\frac{\alpha}{2}\left\<\left\< \bcd_\alpha \bar\vartheta \left[
\bar\varepsilon^{\alpha\beta\gamma\delta} \left(
\bcd_{(\mu}\barh_{|\beta|}{}^\sigma\ \bcd_{\nu)}
\bcd_\delta\barh_{\sigma\gamma} \right. \right.\right.\right.\nn\\
&&\qquad\left.
{}-\bcd^\rho \barh_{\beta(\mu} \bcd_{|\delta|} \bcd_\rho \barh_{\nu)\gamma}
\right) \nn\\
&&\qquad\left.\left.\left.
{}-2 \bar\varepsilon^\alpha{}_{(\mu}{}^{\gamma\delta}
\bcd^\rho\barh_{\nu)\sigma} \bcd_\delta\bcd_\rho \barh^\sigma{}_\gamma
\right]
 \right\>\right\> \ ,
\end{eqnarray}
and
\begin{equation}
\label{eq:CSeffT2}
T_{\CS\mu\nu}^{\eff,2} = -\epsilon^2\alpha\left\<\left\<  \bcd_\sigma\bcd_\rho\bar\vartheta \ 
\bar\varepsilon^\alpha{}_{(\mu}{}^{\gamma\delta} \bcd_{|\alpha|}
\barh_{\nu)}{}^\sigma \bcd_\gamma \barh^\rho{}_\delta \right\>\right\> \ .
\end{equation}
\end{subequations}

\subsection{Imposing the On-Shell Condition}

The equation of motion may be imposed anywhere
$\bar\square\barh_{\alpha\beta}$ may be formed in
$T^\eff_{\CS\mu\nu}$ via integration by parts. 
There is no contraction of derivative operators onto each other in
$T^{\eff,2}_{\CS\mu\nu}$, so it remains unchanged.
In the final two terms
of $T^{\eff,1}_{\CS\mu\nu}$, the derivative operator $\bcd_\rho$ may
be moved onto $\bcd_\alpha\bar\vartheta\
\bcd^\rho\barh_{\kappa\lambda}$. This would generally make two terms,
but the term proportional to $\bcd_\alpha\bar\vartheta\
\bar\square\barh_{\kappa\lambda}$ is $\scO(\zeta_{\GW}^2)$ relative to
the Isaacson piece, so we only keep one term. This gives
\begin{eqnarray}
\label{eq:CSeffT1again}
T_{\CS\mu\nu}^{\eff,1} &=& -\epsilon^2\frac{\alpha}{2}\left\<\left\< 
\bar\varepsilon^{\alpha\beta\gamma\delta} \left(
\bcd_\alpha \bar\vartheta\ 
\bcd_{(\mu}\barh_{|\beta|}{}^\sigma\ \bcd_{\nu)}
\bcd_\delta\barh_{\sigma\gamma} \right.\right.\right.\nn\\
&&\qquad\left.
{}+\bcd_\rho\bcd_\alpha\bar\vartheta\ 
\bcd^\rho\barh_{\beta(\mu} \bcd_{|\delta|}\barh_{\nu)\gamma} \right)\nn\\
&&\qquad\left.\left.{}+2 \bcd_\rho\bcd_\alpha\bar\vartheta\ 
\bar\varepsilon^\alpha{}_{(\mu}{}^{\gamma\delta}\ 
\bcd^\rho\barh_{\nu)\sigma}\bcd_\delta\barh^\sigma{}_\gamma
\right\>\right\> \ .
\end{eqnarray}

Let us now evaluate $T^\eff_{\MT\mu\nu}$ on shell. Since
$T^\eff_{\MT\mu\nu}$ is $\scO((\zeta_{\GW})^0)$, imposing the equation of
motion Eq.~\eqref{eq:EOMCS1atscriagain} will introduce terms of
$\scO(\zeta_{\GW})$, which are kept since they are the same order as
$T^{\eff,1}_{\CS\mu\nu}$ and $T^{\eff,2}_{\CS\mu\nu}$. We can also
impose a gauge condition. We have already imposed the Lorenz gauge
throughout at the level of the action. We may further specialize this
to the TT gauge. While the TT gauge may not be imposed globally, it
may be imposed at $\scri^+$, where the effective stress-energy tensor
is being evaluated. In TT gauge,
\begin{subequations}
\begin{eqnarray}
\label{eq:TMTTT}
T^\eff_{\MT\mu\nu} &=& \epsilon^2\kappa \left\<\left\< \frac{1}{2} \bcd_\mu
\barh_{\alpha\beta} \bcd_\nu \barh^{\alpha\beta}
- \bcd_\rho \barh_{\alpha\mu} \bcd^\rho \barh_\nu{}^\alpha\right.\right.\nn\\
&&\qquad\left.\left.{}-\frac{1}{4} \bar\met_{\mu\nu} \bcd_\rho \barh_{\alpha\beta} \bcd^\rho
\barh^{\alpha\beta} \right\>\right\>\\
&=& T^\eff_{\GR\mu\nu} + T^{\eff,1}_{\MT\mu\nu}+ T^{\eff,2}_{\MT\mu\nu}\,,\nn
\end{eqnarray}
where
\begin{eqnarray}
T^{\eff,1}_{\MT\mu\nu} &=& -\epsilon^2\kappa\left\<\left\< \bcd_\rho
\barh_{\alpha\mu} \bcd^\rho \barh_\nu{}^\alpha \right\>\right\> \\
T^{\eff,2}_{\MT\mu\nu} &=& -\epsilon^2\frac{\kappa}{4}\left\<\left\< 
\bar\met_{\mu\nu}\bcd_\rho
\barh_{\alpha\beta} \bcd^\rho \barh^{\alpha\beta} \right\>\right\>\,.
\end{eqnarray}
Integrating by parts, imposing the equations of motion
Eq.~\eqref{eq:EOMCS1atscriagain}, and integrating by parts again where
appropriate, these contributions to the effective stress-energy tensor
at $\scri^+$ are
\begin{eqnarray}
\label{eq:TMT1again}
T^{\eff,1}_{\MT\mu\nu} &=&-\epsilon^2\left\<\left\<\barh^\alpha{}_{(\mu}
\tilde{T}^{(\vartheta)}_{\nu)\alpha} \right\>\right\> \nn\\
&& {}-\epsilon^2\frac{\alpha}{2} \left\<\left\<
\bcd_\sigma\bcd_\rho\bar\vartheta\ \bcd^\delta \barh^\alpha{}_\mu\ 
\bar\varepsilon^\sigma{}_{\gamma\delta(\alpha} \left(\bcd_{\nu)}
  \barh^{\rho\gamma} - \bcd^\rho\barh_{\nu)}{}^\gamma \right)\right.\right.\nn\\
&&\qquad {}+\left(\mu \leftrightarrow \nu \right) \Big\>\Big\>\,\\
\label{eq:TMT2again}
T^{\eff,2}_{\MT\mu\nu}&=&\frac{1}{4}\bar\met_{\mu\nu}\bar\met^{\alpha\beta}T^{\eff,1}_{\MT\alpha\beta} \,.
\end{eqnarray}
\end{subequations}

Finally, we may write an expression for $T^\eff_{\CS\mu\nu}$ at
$\scri^+$ after imposing the equations of motion,
\begin{subequations}
\begin{eqnarray}
T^\eff_{\CS\mu\nu} &=& T^\eff_{\GR\mu\nu} + \delta T^\eff_{\CS\mu\nu}
\\
\delta T^\eff_{\CS\mu\nu} &=& T^{\eff,1}_{\MT\mu\nu} +
T^{\eff,2}_{\MT\mu\nu} +  T^{\eff,1}_{\CS\mu\nu} +  T^{\eff,2}_{\CS\mu\nu}\,,
\end{eqnarray}
\end{subequations}
where $\delta T^\eff_{\CS\mu\nu}$ contains the Chern-Simons correction
at $\scO(\zeta_{\GW})$. The summands are taken from
Eqs.~\eqref{eq:CSeffT2},~\eqref{eq:CSeffT1again},~\eqref{eq:TMT1again},
and~\eqref{eq:TMT2again}. Putting them together for convenience, the final result is
\begin{widetext}
\begin{eqnarray}
\delta T^\eff_{\CS\mu\nu} &=& -\epsilon^2\left\<\left\<\barh^\alpha{}_{(\mu}
\tilde{T}^{(\vartheta)}_{\nu)\alpha} + \frac{1}{4} \bar\met_{\mu\nu}
\barh^{\alpha\beta}\tilde{T}^{(\vartheta)}_{\alpha\beta}\right\>\right\>
\nn\\
&&{}-\epsilon^2\frac{\alpha}{2} \left\<\left\<
\bcd_\sigma\bcd_\rho\bar\vartheta\ 
\left[
\bcd^\delta \barh^\alpha{}_\mu\ 
\bar\varepsilon^\sigma{}_{\gamma\delta(\alpha} \left(\bcd_{\nu)}
  \barh^{\rho\gamma} - \bcd^\rho\barh_{\nu)}{}^\gamma \right)
+\bcd^\delta \barh^\alpha{}_\nu\ 
\bar\varepsilon^\sigma{}_{\gamma\delta(\alpha} \left(\bcd_{\mu)}
  \barh^{\rho\gamma} - \bcd^\rho\barh_{\mu)}{}^\gamma \right)\right.
\right.\right.\nn\\
&&\qquad\qquad {}+\frac{1}{2}\bar\met_{\mu\nu} 
\bcd^\delta \barh^{\alpha\beta}\ 
\bar\varepsilon^\sigma{}_{\gamma\delta(\alpha} \left(\bcd_{\beta)}
  \barh^{\rho\gamma} - \bcd^\rho\barh_{\beta)}{}^\gamma \right)
+ 2
\bar\varepsilon^\alpha{}_{(\mu}{}^{\gamma\delta} \bcd_{|\alpha|}
\barh_{\nu)}{}^\sigma \bcd_\gamma \barh^\rho{}_\delta
\nn\\
&&\qquad\qquad
\left. {}+ \bar\varepsilon^{\sigma\beta\gamma\delta}
\bcd^\rho\barh_{\beta(\mu} \bcd_{|\delta|}\barh_{\nu)\gamma}
+2 \bar\varepsilon^\sigma{}_{(\mu}{}^{\gamma\delta}\ 
\bcd^\rho\barh_{\nu)\sigma}\bcd_\delta\barh^\sigma{}_\gamma
\right] \nn\\
&&\qquad\qquad \left.\left.{}+ \bar\varepsilon^{\alpha\beta\gamma\delta}
\bcd_\alpha \bar\vartheta\ 
\bcd_{(\mu}\barh_{|\beta|}{}^\sigma\ \bcd_{\nu)}
\bcd_\delta\barh_{\sigma\gamma} \right\>\right\> \,.
\end{eqnarray}
\end{widetext}

In the above, we have organized the terms by their scaling with powers
of wavelength. The first line contains terms which scale as
$\lambda_\GW^0$ and $\lambda_\GW^{-1}$; the first of these corresponds to a
``mass'' term in the effective stress-energy tensor. Both of these
scale more slowly with inverse wavelength than the GR contribution, so
they are subdominant. The next three lines have the same scaling with
inverse wavelength as GR, $\lambda_\GW^{-2}$. The final line scales more
strongly with inverse wavelength, $\lambda_\GW^{-3}$. This term in
principle could dominate over the GR term in the high frequency limit.

Notice that the effective stress-energy tensor presented here is applicable to both the 
dynamical and the non-dynamical version of CS gravity. Also note that if $\vartheta$ were
a constant, rather than a function, the effective stress-energy tensor would be identical to that
of GR (which is expected, since, in that case, the modification to the
action is purely a boundary or topological term).

\subsection{In dynamical CS gravity}
\label{sec:DCSTab}
From asymptotic arguments, we can argue that
$\delta T^\eff_{\CS\mu\nu}$ does not
contribute to dissipation laws at $\scri^+$ in the dynamical version of CS gravity.
As mentioned in \autoref{sec:atscriplus}, the dissipation of energy,
linear and angular momentum of a system is computed by integrating
components of the stress-energy tensor on a 2-sphere at $\scri^+$.
Since the area of the 2-sphere grows as $r^2$, for the dissipation integrals to be finite,
the components of the stress-energy tensor must fall-off at least as
$r^{-2}$. In fact, only the $r^{-2}$ part of the stress-energy
contributes as one takes the $r\to\infty$ limit. Therefore, any part
of the stress-energy tensor that decays faster than $r^{-2}$ does not
contribute to dissipation laws.

The CS correction to the effective stress-energy tensor, $\delta
T^\eff_{\CS\mu\nu}$, always falls off faster than $r^{-2}$ in the
dynamical theory. To see this, we must analyze the behavior of
$\vartheta$, which is restricted. This restriction comes from
demanding that the field $\vartheta$ sourced by an isolated system and
in an asymptotically flat space contains a finite amount of
energy. The energy in $\vartheta$ is computed by integrating the
time-time component of $T^{(\vartheta)}_{\mu \nu}$ on a hypersurface
of constant time and over all space. For the energy to be finite, the
integral $\int^\infty \ (\cd \vartheta)^2 \ r^2 \rmd r$ (in an
asymptotically flat, Cartesian spatial slice, appropriate to
$\scri^{+}$) must be finite. This restricts $\cd\vartheta$ to fall off
at least faster than $r^{-3/2}$.  We then conclude that the CS
correction to the energy-momentum tensor must vanish at $\scri^{+}$,
as $T^{\eff,1}_{\MT\mu\nu}$, $T^{\eff,2}_{\MT\mu\nu}$,
$T^{\eff,1}_{\CS\mu\nu}$ and $T^{\eff,2}_{\CS\mu\nu}$ decay at least
as $r^{-7/2}$ or faster.

The only contribution at $\scri^+$ to the effective stress-energy of
GWs in dynamical CS gravity which decays as $r^{-2}$
is the GR part,
\begin{equation}
T^\eff_{\CS\mu\nu} = T^\eff_{\GR\mu\nu}\,.
\end{equation}

Again, we stress that this only accounts for the outgoing
GW radiation. However, the same argument as in
\autoref{sec:atscriplus} holds; the correction to the energy flux absorbed
by trapped surfaces is only important at the end of an inspiral, both in GR
and deformations away from GR. This is supported by the small
velocity, small mass ratio expansion of~\cite{Mino:1997p287} 
(see also~\cite{Yunes:2009ef,Yunes:2010zj}).

\subsection{In non-dynamical CS gravity}
\label{sec:NDCSTab}
In the dynamical theory, since the scalar field $\vartheta$ must carry
a finite energy, we were able to argue for the vanishing of
$\delta T^{\eff}_{\CS\mu\nu}$ at $\scri^+$. In
the non-dynamical theory, there is no such demand and no further simplification can be
made beyond the vanishing of $T^{(\vartheta)}_{\mu\nu}$.
However, for a particular choice of $\bar\vartheta$ field, the
effective stress-energy tensor may be evaluated. We demonstrate
this below.

\subsubsection*{In the canonical embedding}
The canonical embedding of non-dynamical CS gravity is given by~\cite{jackiw:2003:cmo}
\begin{equation}
v_\mu \equiv \bcd_\mu\bar\vartheta\ \dot{=}\
\left(1/\mu,0,0,0\right)\,,
\end{equation}
in Cartesian coordinates in the asymptotically-flat part of the
spacetime. Approaching infinity, this yields 
\begin{equation}
\bcd_\alpha\bcd_\beta\bar\vartheta = 0
\end{equation}
so by extension $T^{\eff,2}_{\CS\mu\nu} = 0$, the first-order equation
of motion becomes $\bar\square\barh_{\mu\nu} = 0 + \scO(\zeta_{\GW}^2)$, the
final two terms of $T^{\eff,1}_{\CS\mu\nu}$ vanish, and
$T^{\eff,1}_{\MT\mu\nu} = 0 = T^{\eff,2}_{\MT\mu\nu}$. Notice that here there is no
amplitude birefringence in flat spacetime as $\ddot{\vartheta} = 0$~\cite{Alexander:2004wk,Alexander:2007kv,Yunes:2010yf,2010PhRvD..82f4017Y}.

The first term of $T^{\eff,1}_{\CS\mu\nu}$ is the only
$\scO(\zeta_{\GW})$ correction which survives. The total stress-energy
tensor in the canonical embedding of non-dynamical Chern-Simons
gravity at $\scri^+$, with this correction, is
\begin{eqnarray}
T^\eff_{\CS\mu\nu} &=& T^\eff_{\GR\mu\nu} + \delta T^\eff_{\CS\mu\nu}\,,\nn\\
\delta T^\eff_{\CS\mu\nu} &=& - \epsilon^2 \frac{\alpha}{2}
\left\<\left\< \bcd_\alpha\bar\vartheta\
\bar\varepsilon^{\alpha\beta\gamma\delta} \bcd_{(\mu}
\barh_{|\beta|}{}^\sigma \bcd_{\nu)} \bcd_\delta\barh_{\sigma\gamma}
\right\>\right\> \nn\\
&=& + \epsilon^2 \frac{\alpha}{2\mu} \left\<\left\<
\bar\varepsilon^{ijk} \bcd_{(\mu} \barh_{|i|}{}^\sigma
\bcd_{\nu)}\bcd_k\barh_{\sigma j} \right\>\right\>\,,
\end{eqnarray}
where $\bar\varepsilon^{ijk}$ is the Levi-Civita tensor on the 3-space
orthogonal to $(\pd/\pd t)^\mu$, and the sign change arises from the
factor of $\sign(\met)$ in Eq.~\eqref{eq:LeviCivitaDef}.

From the form of the correction $\delta T^\eff_{\CS\mu\nu}$, we can
briefly mention the leading modification to radiation reaction in a
binary inspiral at Newtonian order. At this order, there is no
modification to the trajectories of the two bodies from the
$\bar\vartheta$ field. Since the first-order equation of motion is
identical to that of GR at order $\scO(\zeta_{\GW})$, the leading solution
to $\barh_{\mu\nu}$ is the same as in GR, $\barh_{\mu\nu} =
\barh^\GR_{\mu\nu}$.

Inserting this solution in TT gauge into $\delta T^\eff_{\CS\mu\nu}$,
the energy, linear momentum, and angular momentum radiated by the
system can be computed. Adopting a Cartesian coordinate system at
asymptotic infinity, the correction to the radiated quantities is
given by
\begin{subequations}
\label{eq:radQuantsDef}
\begin{eqnarray}
\delta \dot{E}^\CS &=& -\int \rmd\Omega\ r^2\ \delta
T^\eff_{\CS 0j}n_j =+\int\rmd\Omega\ r^2\ \delta T^\eff_{\CS 00} \\
\delta \dot{P}^\CS_i &=& +\int \rmd\Omega\ r^2\ \delta
T^\eff_{\CS ij}n_j = -\int\rmd\Omega\ r^2\ \delta T^\eff_{\CS i0} \\
\delta \dot{J}^\CS_i &=& -\int \rmd\Omega\ r^2\ 
\varepsilon_{ijk}x_j\delta T^{(-3)}_{\CS kl} n_l\,,
\end{eqnarray}
\end{subequations}
where $\delta T^{(-3)}_{\CS\mu\nu}$ is the part of $\delta T^\eff_{\CS
  \mu\nu}$ which decays as $r^{-3}$~\cite{Thorne:1980rm}. In
evaluating these integrals, the only angular dependence is in factors
of $n_i$ or $x_i$. An angular integral of an odd number of such factors
vanishes, while an integral of an even number of them reduces to a
symmetrized product of Kronecker delta tensors. These factors arise
explicitly in the definitions of Eqs.~\eqref{eq:radQuantsDef} and from
spatial derivatives acting on $\barh_{\mu\nu}$ in
$T^\eff_{\mu\nu}$. The most important difference between
$T^\eff_{\GR\mu\nu}$ and $\delta T^\eff_{\CS\mu\nu}$ is the parity of
the number of derivatives, which leads to the following behaviour.

In GR, the leading contribution to $\dot{E}^\GR$ is from the
(mass quadrupole)${}^2$ combination. Compare this with the same integral
for $\delta T^\eff_{\CS 00}$, where the (mass quadrupole)${}^2$ term has an
odd number of factors of $n_i$, and thus vanishes. The leading
contribution is then from the product of the mass quadrupole and mass
octupole.

The same situation takes place in calculating $\dot{J}_i$. In GR, the
leading contribution is from the product of mass quadrupole with
itself. In the correction from CS gravity, the mass quadrupole squared
term has an odd number of factors of $n_i$; the dominant contribution
is again from the mass quadrupole times the mass octupole.

Finally, the situation is different in the calculation of
$\dot{P}_i$. In GR, the quadrupole squared contribution to $\dot{P}_{i}$ has 
an odd number of $n_i$ factors. The dominant contribution is from the mass quadrupole
times the mass octupole. However, for the CS correction, the
quadrupole squared term has an even number of factors of $n_i$.
Using
\begin{equation}
\barh^\TT_{ij} = \frac{1}{8\pi\kappa r}\ddot{\mathcal{I}}^\TT_{ij}(t-r)\,,
\end{equation}
this evaluates to
\begin{equation}
\delta \dot{P}^\CS_i = -\frac{\alpha}{120 \pi \kappa^2 \mu} 
\varepsilon_{ijk} \mathcal{I}^{(3)}_{lj} \mathcal{I}^{(4)}_{lk}\,,
\end{equation}
where $\mathcal{I}_{ij}$ is the reduced quadrupole moment of the
matter, and $\mathcal{I}^{(n)}_{ij} \equiv (\rmd/\rmd t)^n
\mathcal{I}_{ij}$.

For a binary in a circular orbit about the $\hat{z}$-axis with masses $m_1$,
$m_2$, total mass $m=m_1+m_2$, symmetric mass ratio $\eta =
m_1m_2/m^2$, separation $d$, and orbital frequency $\omega$, we find
the momentum flux correction to be
\begin{subequations}
\begin{equation}
\delta \dot{P}^\CS_z = -\frac{8\alpha}{15\pi\kappa^2\mu}
(\eta m d^2)^2\omega^7\,,
\end{equation}
or, in terms of the velocity $v=\omega d$, with
Kepler's third law $v^{2} = m/d$,
\begin{equation}
\delta \dot{P}^\CS_z = - \frac{128}{15}
\left(\frac{\alpha}{\kappa\mu m}\right) \eta^{2} v^{13}\,,
\end{equation}
\end{subequations}
where notice that the quantity in parentheses is dimensionless.
This is to be compared with the leading momentum luminosity in GR,
which is proportional to $\dot{P}^\GR_z \propto \eta^2 v^{11} \delta m/m$, where
$\delta m = m_1-m_2$~\cite{Blanchet:2005rj}. Although the GR effect is two powers
of $v$ stronger, it depends on the difference in masses, whereas the
non-dynamical CS correction only depends on the total mass. This implies that in the
limit of comparable masses $m_{1} \approx m_{2}$, the recoil velocity would not
asymptote to zero in CS gravity, as it does in GR for non-spinning binaries.

A physical interpretation of this effect is related to the
parity-violating nature of the theory. When one chooses a canonical
embedding, the action becomes parity-violating as the Pontryagin
density is parity odd. The embedding coordinate chooses a (temporal)
direction to which the modification to the Einstein equations can
couple to, inducing a new term in the stress-energy that is
proportional to the curl of the metric perturbation. Because kicks are
predominantly generated during merger, the CS modification is indeed
dominant over the GR result, leading to the first, non-linear,
strong-field modification computed in CS gravity.

\section{Effective Stress-Energy Tensor of Modified Gravity Theories}
\label{sec:gen-alt-theories}

Let us now consider a broader class of modified gravity
theories. There is an infinite variety of GR modifications one
could construct. However, there are several
properties that are desirable and that we require here:
\begin{enumerate}
\item Metric theories: the action depends on a symmetric metric tensor that controls the spacetime dynamics.
\item Deformations of GR: analytically controllable and small corrections to the Einstein-Hilbert action with a continuous GR limit.
\item High-Rank Curvature: corrections depend on quadratic or higher products of the Riemann tensor, Ricci tensor, or Ricci scalar.
\item Minkowski stable: the theory must admit Minkowski spacetime as a
      stable vacuum solution, and future null infinity should be
      asymptotically flat for isolated matter spacetimes.
\end{enumerate}

Besides the metric, there may be new fields introduced which are
considered part of the ``gravity sector''. This distinction means that
said fields are not minimally coupled, i.e.~they may be coupled to
connection and curvature quantities. These additional fields may be
of any spin: scalars, spinors, vectors, etc. For simplicity, we will
only consider scalar fields here, but the results may also be extended to
higher spin fields. Scalar fields are well-motivated from quantum completions
of GR, e.g.~moduli fields are common appearances in string theoretical 
models~\cite{Polchinski:1998rr}. 

\subsection{Action}
In defining a modified gravity theory, let us consider what terms may
arise in the action. These terms must include the
Einstein-Hilbert and matter terms, along with modifications built from 
additional scalar fields and curvature invariants. Additionally, it
ought to contain a dynamical term for the scalars that couple to the
curvature invariants, as we will motivate in Sec.~\ref{sec:alt-dynamical}.

In principle, there are an infinite number of curvature invariants to
consider. The first few of these are simple to construct: $\Lambda,\
R,\ R^2, \ \cd_\mu R \cd^\mu R,\ R_{\mu\nu}R^{\mu\nu},\
R_{\alpha\beta\mu\nu}R^{\alpha\beta\mu\nu},\ \ldots$, where $\Lambda$ is
any scalar constant, e.g.~the cosmological constant. These may be
specified by their rank, $r$, which is the number of curvature tensors
which are contracted together, and by further specifying a list of $r$
non-negative integers $\{\lambda_1,\ldots,\lambda_r\}$, where
$\lambda_i$ specifies the number of derivatives acting on the
$i^\mathrm{th}$ curvature tensor. For a rank $r$ and case
$\{\lambda_i\}_{i=1}^r$, there are a finite number of independent
curvature invariants corresponding to the number of ways to contract
indices.
Thus all curvature invariants may be countably
enumerated, assigning some number $n$ to each independent
invariant. 

We here consider only combinations of {\emph{algebraic}} curvature invariants, 
i.e.~$\lambda_i = 0$ for all cases. This means we do not allow modifications that
depend on derivatives of curvature tensors. Such a simplification is a good one, 
from the standpoint that it automatically guarantees the field equations to be no higher
than fourth-order.

Consider then a modified gravity theory defined by the action
\begin{subequations}
\begin{equation}
\label{eq:gen-alt-action}
S = S_\EH + S_\MAT + S_\INT + S_\vartheta\,,
\end{equation}
where $S_{\vartheta}$ is the canonical kinetic term for $\vartheta$,
\begin{equation}
S_\vartheta = -\frac{\beta}{2} \int \rmd^4 x \sqrt{-\met}\ \met^{\mu\nu}
\left[ (\cd_\mu\vartheta)(\cd_\nu\vartheta) + 2 V(\vartheta) \right]\,,
\end{equation}
with $V$ an arbitrary potential function; and
where $S_\INT$ is the interaction term between the scalar $\vartheta$
and some algebraic combination of curvature tensors, for example
\begin{eqnarray}
S_{\INT,0} &=& \alpha_0\int \rmd^4x \sqrt{-\met}\ f_0(\vartheta) \Lambda\\ 
S_{\INT,1} &=& \alpha_1\int \rmd^4x \sqrt{-\met}\ f_1(\vartheta) R\\ 
S_{\INT,2} &=& \alpha_2\int \rmd^4x \sqrt{-\met}\ f_2(\vartheta) R^2\,,
\end{eqnarray}
or generally 
\begin{eqnarray}
S_\INT &=& \alpha\int \rmd^4x \sqrt{-\met}\ f(\vartheta) \scR\,,
\label{eq-interactionS}
\end{eqnarray}
with $f$ an arbitrary ``coupling function'' and $\scR$ an algebraic combination of
curvature invariants. Alternatively, notice that we could have assigned each term
proportional to $\alpha_{i}$ a separate $\vartheta_{i}$ coupling with its associated
kinetic and potential terms. The arguments presented below would also hold for such
constructions.
\end{subequations}

\subsection{Dynamical scalar fields}
\label{sec:alt-dynamical}
The requirement for the scalar $\vartheta$ to be dynamical arises
from demanding diffeomorphism invariance in the theory. Consider the
infinitesimal transformation of the action under a diffeomorphism
generated by the vector field $v^\mu$. Specifically, look at the terms
containing $\vartheta$, i.e.~the sum $S_{\mbox{\tiny mod}} = S_\INT+S_\vartheta$. The
infinitesimal transformation under the diffeomorphism is
\begin{multline}
\delta S_{\mbox{\tiny mod}} = \int\rmd^4x \left( \frac{\delta}{\delta \met_{\mu\nu}}
\scL_\INT \right) \scL_v \met_{\mu\nu}
+ \left(\frac{\delta}{\delta\vartheta}
\scL_\INT \right) \scL_v \vartheta \\
+\int\rmd^4x \left(\frac{\delta}{\delta \met_{\mu\nu} }
\scL_\vartheta \right) \scL_v \met_{\mu\nu}
+ \left(\frac{\delta}{\delta \vartheta}
\scL_\vartheta \right) \scL_v \vartheta \,,
\end{multline}
where $\scL_\INT$ is the interaction Lagrangian density, $\scL_\vartheta$
is the kinetic Lagrangian density, and $\scL_v$ stands for the Lie derivative
along $v^\mu$.

For a theory to be diffeomorphism invariant, the infinitesimal
transformation in the total action must vanish, $\delta S = 0$. Since
$\scL_v \vartheta$ may be arbitrary for some $\vartheta$ and some
$v^\mu$, the functional multiplying $\scL_v\vartheta$ must vanish
for $\delta S$ to vanish. This means
\begin{equation}
\label{eq:diffeocondition}
\frac{\delta}{\delta \vartheta} \left( \scL_\INT + \scL_\vartheta
\right) = 0\,.
\end{equation}
When the scalar field $\vartheta$ has dynamics, i.e.~$\beta\ne 0$,
then Eq.~\eqref{eq:diffeocondition} is identical to the equations of
motion of the field $\vartheta$ and is therefore automatically
satisfied. However, if the field is \emph{not} dynamical, $\beta =
0$, then Eq.~\eqref{eq:diffeocondition} gives
\begin{equation}
\label{eq:nondynamicalcondition}
f^\prime(\vartheta) \scR = 0\,.
\end{equation}
Except in the case where $f^\prime(\vartheta) = 0$, this is an
additional constraint on the geometry of spacetime, namely that
$\scR = 0$. Given that the equations of motion already
saturate the number of equations for the degrees of freedom present,
this would be an overconstrained system. This is in fact the case in the
non-dynamical version of CS gravity, as discussed in~\cite{Grumiller:2007rv,Yunes:2007ss}. 
We therefore only admit dynamical scalar fields, or terms with no scalar 
field dependence ($f(\vartheta)=const.$).

\subsection{Special cases: zeroth and first rank}
Before doing a calculation for a general curvature invariant
$\scR$, let us briefly discuss some special cases. As we will see,
curvature invariants of zeroth and first rank will not be
considered.

\subsubsection*{Zeroth rank}
\label{sec:alt-zeroth-order} 
At zeroth rank, there is only one algebraic curvature invariant: a
constant.
The non-constant part of $f_0(\vartheta)$ may simply be reabsorbed
into the potential $V(\vartheta)$. This gives a minimally coupled
scalar field, which may be absorbed into $S_\MAT$. The constant part
of $f_0$ leads to a ``cosmological constant.'' Since we are only
considering theories which are Minkowski stable and asymptotically
flat, this cosmological part must vanish.

\subsubsection*{First rank}
\label{sec:alt-first-order} 
There is only one algebraic curvature invariant of rank 1, the Ricci
scalar $R$. If we allow $f$ to be a non-constant function, we would
have a classical scalar-tensor theory, akin to Brans-Dicke theory (see e.g.~\cite{lrr-2006-3}). 
The effective GW stress-energy tensor for scalar-tensor theory has been
computed, for example in Brans-Dicke theory (see~\cite{1993tegp.book.....W}), 
so we do not consider it here. Since we already include the Einstein-Hilbert term in 
Eq.~\eqref{eq:gen-alt-action}, there can be no additional term linear
in $R$ without affecting Newton's constant. Thus we only consider
quadratic and higher rank curvature invariants.

\subsection{Cubic and Higher Ranks}
\label{sec:alt-higher-order}
At cubic rank, one can easily show that there are five algebraic invariants that may not be 
factored as products of lower rank invariants 
($R^\mu{}_\nu R^\nu{}_\rho R^\rho{}_\mu,\ R^{\mu\nu}R^{\alpha\beta}  R_{\mu\alpha\nu\beta}$,
$R^{\alpha\beta}{}_{\mu\nu} R^{\mu\nu}{}_{\rho\sigma} R^{\rho\sigma}{}_{\alpha\beta}$,
${}^{*}R^{\rho\sigma\mu\nu} R^{\kappa\lambda}{}_{\mu\nu} 
R_{\rho\sigma\kappa\lambda}$, and ${}^{*}R_{\rho\sigma}{}^{\alpha \beta} 
R^\rho{}_\alpha R^\sigma{}_\beta$) and four that may be factorized 
($R^3$, $R R_{\mu\nu} R^{\mu\nu}$, $R R_{\mu\nu\alpha\beta}
R^{\mu\nu\alpha\beta}$, and $R\,{}^{*}R^\rho{}_{\sigma}{}^{\mu \nu} 
R^\sigma{}_{\rho\mu\nu}$). The arguments
that follow work for all of them, so for concreteness we choose just
one: $R^\mu{}_\nu R^\nu{}_\rho R^\rho{}_\mu$. The modification to the
action arising from this term is
\begin{equation}
\label{eq:cubicExampleAction}
S_{\mbox{\tiny{ex.5}}} = \alpha\int\rmd^4x \sqrt{-\met}\  f(\vartheta)
R^\mu{}_\nu R^\nu{}_\rho R^\rho{}_\mu\,.
\end{equation}
The contribution from this term to the effective action at second
order is
\begin{multline}
\label{eq:cubicExampleEff}
S^{\eff (2)}_{\mbox{\tiny{ex.5}}} = \epsilon^2 \alpha \int\rmd^4 x
\sqrt{-\bar\met} \left[ \frac{\tilde{h}}{2}
\left( f^\prime(\bar\vartheta)\tildetheta \ \bar{R}^\mu{}_\nu
\bar{R}^\nu{}_\rho \bar{R}^\rho{}_\mu \right.\right.\\
\left. + 3 f(\bar\vartheta) \ \tilde{R}^\mu{}_\nu
\bar{R}^\nu{}_\rho \bar{R}^\rho{}_\mu\right)
+ 3 f^\prime(\bar\vartheta) \tildetheta \ \tilde{R}^\mu{}_\nu
\bar{R}^\nu{}_\rho \bar{R}^\rho{}_\mu \\
\left. + 3 f(\bar\vartheta) \tilde{R}^\mu{}_\nu
\tilde{R}^\nu{}_\rho \bar{R}^\rho{}_\mu\right] \,.
\end{multline}
Immediately we see that all terms have at least one power of
background curvature tensors. This means that each term can be written
similarly to an earlier example in Sec.~\ref{sec:atscriplus}, in
Eq.~\eqref{eq:curvatureExample}. When evaluating this effective stress-
energy tensor at $\scri^+$, all of the background curvature tensors
vanish. This automatically implies that cubic and higher rank terms in the action
do not contribute to the effective stress energy tensor at asymptotically-flat, 
future null infinity.

The stress-energy tensor is then given by the MacCallum-Taub tensor, Eq.~\eqref{eq:TMT}, 
which need not be identical to the GR one yet, as one must first impose the first-order
equations of motion at $\scri^+$. These equations could be modified by the
introduction of higher-order operators in the action. Let us analyze such equations
again through the example of Eq.~\eqref{eq:cubicExampleAction}. As the 
calculation depends only on the rank of the curvature invariant
appearing in the action and not on its specific form, the results shown below
extend to all cubic and higher rank algebraic curvature invariants as well.

The equation of motion arising from Eq.~\eqref{eq:cubicExampleAction}
is
\begin{multline}
\label{eq:cubicExampleEOM0}
\kappa G_{\mu\nu} + 3\alpha f(\vartheta) R_{\mu\beta}
R^\beta{}_\gamma R^\gamma{}_\nu - \frac{\alpha}{2}\met_{\mu\nu}
f(\vartheta) R^\alpha{}_\beta R^\beta{}_\gamma R^\gamma{}_\alpha \\
+ \frac{3\alpha}{2} \left[ \met_{\mu\nu} \cd_\alpha\cd_\beta \left(
    f(\vartheta) R^\alpha{}_\gamma R^{\gamma\beta} \right) + \square \left( f(\vartheta)
    R_{\mu\gamma} R^\gamma{}_\nu \right) \right.\\
\left. -2\cd_\beta \cd_{(\mu} \left( f(\vartheta) R^\gamma{}_{\nu)}
    R^\beta{}_\gamma \right)\right] = T^\MAT_{\mu\nu} +
T^{(\vartheta)}_{\mu\nu} \,.
\end{multline}
The important feature to note is that all terms containing $\alpha$,
that is, all terms deforming away from GR, are cubic or quadratic in
curvature tensors. This is a general feature: from a term in the
action of rank $r$, terms in the equations of motion will be of rank
$r$ and rank $r-1$.

Now consider evaluating the first-order equations of motion at
asymptotically-flat, future null infinity, which we need in order to
put the MacCallum-Taub stress-energy tensor on-shell. We will not
write out the full first-order equations of motion; it suffices to say
that the modification terms (those terms containing $\alpha$) are of
rank $r$, $r-1$, and $r-2$ in the first-order equations of
motion. When going to $\scri^+$, only the terms of rank 0 survive,
e.g.~$\bar\square\barh_{\mu\nu}$.

Immediately we see that the only modifications to the action that
affect the first-order equations of motion at $\scri^+$ are those of
rank 2 and lower. Thus for modifications that are cubic and higher,
the first-order equations of motion at $\scri^+$ are simply those of
GR, $\bar\square\barh_{\mu\nu} = 0$.

Inserting this asymptotically-flat, on-shell condition into the
MacCallum-Taub stress-energy tensor yields the Isaacson stress-energy
tensor. Cubic and higher rank modifications to the Lagrangian do not
modify the effective stress-energy tensor due to GWs. 
We again emphasize that radiation reaction will still be
different in a higher order theory because of different motion in the
strong field, additional energy carried in the scalar field
$\vartheta$, and energy carried down horizons being different. But the
energy of a GW at $\scri^+$ is the same as in GR.

\subsection{Quadratic terms}
\label{sec:alt-second-order}

Let us now consider quadratic deformations to the action, as these are the
only ones left to study, and let us classify the types of modifications possible.
There are two important characteristics that we use for such a classification. 
The first depends on the nature of the curvature
quantity $\scR$. This quantity may either be topological or not. A
curvature quantity that is topological may be expressed as the
divergence of a current, $\scR = \cd_\mu \mathcal{K}^\mu$. As
we mentioned in Sec.~\ref{sec:CS}, Eq.~\eqref{eq:pontdiv}, the
Pontryagin density, $\pont$, is a topological curvature
invariant. In metric gravity, the only other non-vanishing, algebraic, second
rank curvature invariant that is topological is the Gauss-Bonnet
term,
\begin{equation}
\label{eq:gauss-bonnet}
\mathcal{G} = R_{\alpha\beta\mu\nu}R^{\alpha\beta\mu\nu}
- 4 R_{\mu\nu} R^{\mu\nu} + R^2 \,,
\end{equation}
as the Nieh-Yan invariant vanishes in torsion-free theories.

The second characteristic we can use to classify theories is 
the behaviour of the scalar field $\vartheta$, which depends on the
potential $V(\vartheta)$. The two possibilities are a potential that
is flat, $V(\vartheta)=0$, or one that is non-flat, $V$ varying with
$\vartheta$. A flat potential does not choose out any preferred values
of the scalar field, whereas a non-flat potential must be bounded from
below for stability, and thus has a global minimum (or several
minima). The presence or absence of a preferred field value is
important in the limit going to $\scri^+$.

For non-flat potentials, without loss of generality, the global
minimum can be shifted to $\vartheta=0$ by simultaneously shifting the
potential function and the coupling function $f$. Such a shift does not
affect the derivative term in the kinetic term for $\vartheta$, since
it simply adds a global constant to the field. The only problematic
situation is if the global minimum is in the limit $\vartheta \to \pm
\infty$, which we do not allow here.

We begin by discussing the asymptotic behaviour of $\vartheta$,
which satisfies the sourced wave equation
\begin{subequations}
\begin{equation}
\label{eq:EOMvarthetan}
\beta \left( \square \vartheta - V^\prime(\vartheta)
\right) = -\alpha f^\prime(\vartheta) \scR \,.
\end{equation}
At $\scri^+$, the right hand side vanishes. Furthermore, if we are
interested in static or quasistatic background solutions for
$\bar\vartheta$ around which we can expand, time derivatives in the
d'Alembertian will vanish leaving only the Laplacian,
\begin{equation}
\label{eq:EOMvarthetanatscri}
\bcd^2 \bar\vartheta - V^\prime(\bar\vartheta) = 0 \,.
\end{equation}
For a non-flat potential, $V^\prime(\vartheta)\ne 0$, the zeroth order
asymptotic solution will be $\bar\vartheta$ going to the minimum of
the potential, which we have shifted to $\bar\vartheta = 0$.

For a flat potential, the background equation
of motion for $\vartheta$, Eq.~\eqref{eq:EOMvarthetanatscri}, at
$\scri^+$ becomes
\begin{equation}
\label{eq:EOMvarthetanflat}
\bcd^2\bar\vartheta = 0 \,.
\end{equation}
\end{subequations}
There are two asymptotic solutions: $\vartheta$ asymptotes to a constant or
$\vartheta$ asymptotes to a function linear in Cartesian coordinates. The latter
case would contribute a constant stress-energy tensor
$T^{(\vartheta)}_{\mu\nu}$ at $\scri^+$. This would lead to an
asymptotically de~Sitter spacetime, not an asymptotically flat
spacetime. Therefore, we only consider the case where $\vartheta$ asymptotes 
to a constant.

The equation of motion Eq.~\eqref{eq:EOMvarthetanflat} does not
determine to what value $\bar\vartheta$ asymptotes. A boundary
condition is required in this case. Again, without loss of generality,
for some given asymptotic value determined by some boundary condition, the
field and coupling function $f(\vartheta)$ may be shifted so as to
redefine the asymptotic value to be $\vartheta\to 0$ without changing
the physics.

A boundary condition is \emph{not} required if the theory is ``shift
symmetric.'' In a shift symmetric theory, the translation operation 
$\vartheta \to \vartheta + c$, where $c$ is a constant, leaves the equations of motion 
invariant. Such a theory, therefore, must have equations of motion that depend only
on the derivative $\cd_\mu \vartheta$. Such is the case, for example,
if the action depends on a topological term multiplied by some scalar field, 
$f(\vartheta) \cd_\mu \mathcal{K}^\mu$, as then the action can be rewritten as  
$(\cd_\mu f(\vartheta)) \mathcal{K}^\mu$ via integration by parts. Of course, in this case,
the potential must also be flat and $f$ must be linear for the theory to be
shift-symmetric. Such types of corrections arise naturally in the low-energy limit of string theory~\cite{alexander:2004:lfg,Alexander:2004xd,2009PhRvD..80f5003G,Alexander:2009tp}.

Let us rewrite the action and split the interaction
term into a dynamical and non-dynamical part. Since we can always
shift the field, potential, and coupling function so that the
asymptotic value is $\vartheta\to 0$, let us define
\begin{subequations}
\begin{eqnarray}
\alpha^\prime & \equiv & \alpha f(0)\\
F(\vartheta)   & \equiv & f(\vartheta) - f(0) \,.
\end{eqnarray}
Then, the interaction term in Eq.~\eqref{eq-interactionS} may be rewritten as
\begin{eqnarray}
\label{eq:int-split}
S_\INT &=& S_{\mbox{\tiny n-d}} + S_\DYN \\ \nn
&=& \alpha^\prime \int \rmd^4x \sqrt{-\met} \ \scR + \alpha \int
\rmd^4x \sqrt{-\met} \ F(\vartheta) \scR \,.
\end{eqnarray}
\end{subequations}
The first term is the non-dynamical part, i.e.~the part that does not
couple to the scalar field, while the second part is the dynamical
part. If $\scR$ is a topological curvature invariant, then the first
term in Eq.~\eqref{eq:int-split} does not contribute to the equations
of motion, as it is the integral of a total derivative.

\subsubsection{Dynamical contribution}
\label{sec:gen-alt-dyn}
Let us perturb $S_\DYN$ to second order to calculate the contribution to the
effective action, keeping in mind that $\tilde{\tilde{\vartheta}}$ and
$\tilde{\tilde{h}}$ do not contribute. This part of the effective
action is
\begin{eqnarray}
\label{eq:dyn-eff}
S_\DYN^{\eff(2)} &=& \epsilon^2 \alpha\int\rmd^4x \sqrt{-\bar\met} \, \left[ 
F(\bar\vartheta)\ \tilde{\tilde{\scR}} \right.\nn\\
&&\qquad{}+ \left(
  \frac{\tilde{h}}{2} F(\bar\vartheta) + F^\prime(\bar\vartheta)\
  \tildetheta \right) \tilde{\scR}  \nn\\
&&\qquad {}+ \frac{1}{8} \left( \tilde{h}^2 -
    2\tilde{h}^{\mu\nu}\tilde{h}_{\mu\nu} \right) F(\bar\vartheta)
 \ \bar\scR \nn\\
&&\qquad \left. {}+ \frac{1}{2} \left( \tilde{h}\ F^\prime(\bar\vartheta)
   \tildetheta + F^{\prime\prime}(\bar\vartheta)\ \tildetheta^2
 \right) \bar\scR \right] \,.
\end{eqnarray}
To determine the contribution to the effective stress energy tensor at
$\scri^+$, again analyze the asymptotic form of all of the fields
appearing in Eq.~\eqref{eq:dyn-eff}. The asymptotic forms are
summarized in Table \ref{tab:asymptotics}.

\begin{table}[tp]
\begin{tabular}{ccccc}
\hline\noalign{\smallskip}
Field & Asymptotic form && Field & Asymptotic form\\
\noalign{\smallskip}\cline{1-2}\cline{4-5}\noalign{\smallskip}
$\bar\met$                                                  & $\scO(1 + r^{-1})$&& $\bcd\bar\vartheta$ & At least $r^{-3/2}$\\
$\bar\Gamma$                                             & $r^{-2}$                 && $\bar\vartheta$ & At least $r^{-1/2}$\\
$\bar{R}$                                                      & $r^{-3}$                && $F(\bar\vartheta)$ & At least $r^{-1/2}$\\
$\bar\scR\sim\bar{R}^2$                              & $r^{-6}$                && $F^{(n)}(\bar\vartheta)$ & $\scO(1)$\\ 
$\tilde{h},\ \bcd^{(n)}\tilde{h}$                       & $r^{-1}$               && $\tildetheta,\ \bcd^{(n)}\tildetheta$ & $r^{-1}$\\
$\tilde{\scR}\sim\bar{R}(\bcd^2\tilde{h})$    & $r^{-4}$                && $\tilde{\tilde{\scR}}\sim(\bcd^2\tilde{h})^2$ & $r^{-2}$     \\
\noalign{\smallskip}\hline
\end{tabular}
\caption{The asymptotic forms of fields appearing in the effective
  action for a rank 2 modification to the action. All tensor indices
  have been suppressed.}
\label{tab:asymptotics}
\end{table}

The simplest way to see that the dynamical part of
the effective action does not contribute at $\scri^{+}$ is to examine the asymptotics of 
$\bar\scR,\ \tilde{\scR}$, and $\tilde{\tilde{\scR}}$. Since curvature tensors
$\bar{R}_{\alpha\beta\mu\nu}$ are tidal tensors, they goes as
$r^{-3}$; since $\bar\scR$ contains two curvature tensors, it scales
as $r^{-6}$. The slowest decaying (i.e.~leading) part of
$\tilde{\scR}$ roughly comes from $\bar{R}\tilde{R}$ (with indices
suppressed); the leading part of $\tilde{R}$ is $\bcd^2\tilde{h}$,
which, being radiative, goes as $r^{-1}$. This means that
$\tilde{\scR}\sim r^{-4}$. Similarly, the leading part of
$\tilde{\tilde{\scR}}$ goes as $(\bcd^2\tilde{h}) (\bcd^2\tilde{h})$,
so $\tilde{\tilde{\scR}}\sim r^{-2}$.

Examining the effective action, Eq.~\eqref{eq:dyn-eff}, we see that
there are no terms that decay as $r^{-2}$, which are the only ones
that can contribute to the GW effective stress-energy tensor at
$\scri^+$. Any term with $\bar\scR$ or $\tilde{\scR}$ already decay
too quickly; only terms with $\tilde{\tilde{\scR}}$ could remain, and
only if they were multiplied by terms that asymptote as $\scO(1)$.

Having performed the splitting into the dynamical and non-dynamical
parts, $\tilde{\tilde{\scR}}$ is multiplied by $\sqrt{-\bar\met}
F(\bar\vartheta)$ in the effective action. This splitting was
specifically constructed so that $F(0) = 0$. Since $F$ must be
differentiable at $\vartheta=0$, $F(\bar\vartheta)$ must go to zero
at least as fast as $\bar\vartheta$ goes to zero, which is at least
$r^{-1/2}$.

We have thus shown that the dynamical part of the interaction term does
not contributed to the effective stress-energy tensor at $\scri^{+}$ directy. 
However, it could still contribute indirectly through the imposition of the first
order field equations. We examine this in a later section. 

\subsubsection{Non-dynamical contribution}
\label{sec:gen-alt-non-dyn}
Let us now consider $S_{\mbox{\tiny n-d}}$ in Eq.~\eqref{eq:int-split}. This term generically
contributes to the effective stress-energy tensor of GWs at
$\scri^+$. To show this contribution, consider the general rank 2
modification as the linear combination of the four independent rank 2
curvature invariants
\begin{subequations}
\begin{equation}
\label{eq:gen-rank-2}
\alpha\scR \equiv \alpha_1 R^2 + \alpha_2 R_{\mu\nu} R^{\mu\nu} +
\alpha_3 R_{\alpha\beta\mu\nu}R^{\alpha\beta\mu\nu} + \alpha_4 \pont
\, ,
\end{equation}
and absorb $f(0)$ into the coefficients $\alpha^\prime_i$ in the
non-dynamical part,
\begin{equation}
\alpha^\prime\scR \equiv \alpha^\prime_1 R^2 + \alpha^\prime_2 R_{\mu\nu} R^{\mu\nu} +
\alpha^\prime_3 R_{\alpha\beta\mu\nu}R^{\alpha\beta\mu\nu} + \alpha^\prime_4 \pont
\, .
\end{equation}
\end{subequations}
Note that this form also includes the Weyl squared invariant, which is a
dependent linear combination of the above terms,
$C^{\alpha\beta\mu\nu}C_{\alpha\beta\mu\nu} = R^2/3 -2 R^{\alpha\beta}
R_{\alpha\beta}+R^{\alpha\beta\mu\nu}R_{\alpha\beta\mu\nu}$, which is
considered in~\cite{Nelson:2010rt,Nelson:2010ru}.
The Pontryagin density $\pont$, being a topological invariant, does
not contribute to the action in $S_{\mbox{\tiny{n-d}}}$, so we may
drop the final term. Similarly, if the linear combination is
proportional to the Gauss-Bonnet (or Euler) invariant, which has
$\alpha_1=1=\alpha_3$, $\alpha_2=-4$, then $\scR$ would be topological
and there would be no contribution to $S_{\mbox{\tiny{n-d}}}$ and
hence no contribution to the effective stress-energy tensor of GWs.

The calculation of the effective action for the non-dynamical term is
straightforward but long, so we do not show the steps here. An outline
of the calculation is to perturb $\sqrt{-\met} \scR$ to second order;
the only parts that may contribute to an effective stress-energy
tensor at $\scri^+$ are of the form $\sqrt{-\bar\met}
\tilde{R}\tilde{R}$, where again we have suppressed indices on the perturbed 
curvature tensor $\tilde{R}$. This is calculated in terms of the trace-reversed
metric perturbation $\barh^{\mu\nu}$. As before, Lorenz gauge may be
imposed at the level of the action. All terms that remain will be of
the form $\bcd_\alpha\bcd_\beta\barh^{\mu\nu}\
\bcd_\kappa\bcd_\lambda\barh^{\rho\sigma}$ with all indices contracted to form a
scalar. If any derivative is contracted onto $\barh^{\mu\nu}$, by
integrating by parts and commuting covariant derivatives, one may form
the Lorenz gauge condition $\bcd_\mu\barh^{\mu\nu}=0$ and ignore the
term in the effective action. Thus the only surviving terms have
derivatives contracted together, which can be put into one of two
forms, $\bar\square\barh^{\mu\nu}\ \bar\square\barh_{\mu\nu}$ and
$\bar\square\barh\ \bar\square\barh$. After the explicit calculation, the
prefactors are found and
\begin{multline}
\label{eq:non-dyn-eff}
S_{\mbox{\tiny{n-d}}}^{\eff(2)} =\frac{\epsilon^2}{4} \int\rmd^4x
\sqrt{-\bar\met}\ \left[ \left(\alpha^\prime_1 - 
\alpha^\prime_3 \right) \bar\square\barh\ \bar\square\barh \right.\\
\left. + \left(\alpha^\prime_2 + 4\alpha^\prime_3 \right) \bar\square
  \barh^{\mu\nu}\ \bar\square\barh_{\mu\nu}\right]\ .
\end{multline}
Note again that $\alpha_4$ does not appear, and if $\alpha_1,\
\alpha_2,\ \alpha_3$ are in the Gauss-Bonnet ratio, then the effective
action of Eq.~\eqref{eq:non-dyn-eff} vanishes.

Putting all indices in their natural positions, so as to expose
implicit metric dependence, and varying the effective action of
Eq.~\eqref{eq:non-dyn-eff} with respect to $\bar\met^{\mu\nu}$, the
contribution to the effective stress-energy tensor is
\begin{eqnarray}
\label{eq:non-dyn-ESET}
T^\eff_{\mbox{\tiny{n-d}}\mu\nu} &=& \epsilon^2 \left\<\left\< \left( \alpha^\prime_1 -
\alpha^\prime_3 \right) \bar\square\barh \left(\bar\square\barh_{\mu\nu} -
\bcd_\mu\bcd_\nu\barh \right) \nn\right.\right.\\
&&\quad+ \left( \alpha^\prime_2 + 4\alpha^\prime_3\right) \left( \bar\square
  \barh_{\alpha\mu}\ \bar\square\barh_\nu{}^\alpha \right.\nn\\
 &&\qquad\left.\left.\left. {}-\bar\square\barh_{\alpha\beta}\
       \bcd_{(\mu}\bcd_{\nu)}\barh^{\alpha\beta}
     \right)\right\>\right\>\, .
\end{eqnarray}

We then find that the effective stress-energy tensor of GWs at
$\scri^+$ is given by the MacCallum-Taub stress-energy tensor 
(coming from the Einstein-Hilbert action) plus the direct contribution from the
non-dynamical part of the rank 2 interaction term,
$T^\eff_{\mbox{\tiny n-d}\mu\nu}$,
\begin{equation}
T^\eff_{\mu\nu} = T^\eff_{\MT\mu\nu} + T^\eff_{\mbox{\tiny n-d}\mu\nu}\,.
\end{equation}
The only remaining part of the calculation is to put the stress-energy tensor
on-shell, that is, to impose the first-order equations of motion at
$\scri^+$.

\subsubsection{First-order equations of motion}
\label{sec:gen-alt-eom1}
We need the first-order equations of motion at $\scri^+$ of the full
theory, including both the dynamical and non-dynamical terms of the
action. At $\scri^{+}$, however, $\bar\vartheta$ 
and $\bcd_\mu \bar\vartheta$ decay at least as $r^{-1/2}$ and
$r^{-3/2}$. The only remaining dependence on $\bar\vartheta$ is
through $f(\bar\vartheta) \to f(0)$.

The general zeroth and first-order equations of motion are quite long, so we
do not reproduce them here, but they do simplify as $r\to\infty$. These
equations are linear in $\barh_{\mu\nu}$ and $\tildetheta$, which are
both radiative and decay as $r^{-1}$. This $r^{-1}$ scaling is the
asymptotic scaling of the first-order equations of motion, as there
are terms of the form $\bar\square\barh_{\mu\nu}$ that appear with no
curvature tensors or background scalar field $\bar\vartheta$
multiplying them. However, all terms containing $\tildetheta$ have
curvature tensors multiplying them, so they decay faster than the
leading behaviour of $r^{-1}$.

Keeping only the terms that go as $r^{-1}$, the first-order equation
of motion in Lorenz gauge at asymptotically-flat, future null infinity is
\begin{subequations}
\label{eq:rank2EOM1atscri}
\begin{eqnarray}
\label{eq:rank2EOM1tensatscri}
\kappa \bar\square\barh_{\mu\nu} &=&  - \left( 2\alpha_1 + \alpha_2 + 2 \alpha_3 \right) f(\bar\vartheta)
\left( \bcd_\mu\bcd_\nu\bar\square - \bar\met_{\mu\nu}
  \bar\square\bar\square \right) \barh \nn\\
&& -\left( \alpha_2 + 4\alpha_3
\right) f(\bar\vartheta) \bar\square\bar\square\barh_{\mu\nu} \,,
\end{eqnarray}
and the trace of this equation is
\begin{equation}
\label{eq:rank2EOM1traceatscri}
\kappa \bar\square\barh = 2\left( 3\alpha_1 + \alpha_2 + \alpha_3
\right) f(\bar\vartheta) \bar\square\bar\square\barh\,.
\end{equation}
\end{subequations}
Again we see that if the $\alpha$ coefficients are in the Gauss-Bonnet
ratio, the GR equation of motion is recovered at $\scri^+$.

This wave equation can be seen to be a massive wave equation for the
auxiliary variable $\tilde{r}_{\mu\nu}\equiv \bar\square\barh_{\mu\nu}$, with mass 
$m \sim 1/\bar{\lambda}$, where
$\bar{\lambda}^2 \sim |\alpha_i| f(0)/\kappa$.
In the weak coupling limit, $\bar\lambda/\lambda_{\GW} \ll 1$, the equations 
simplify considerably. This simplification comes from treating the
solution to the full theory as a {\emph{deformation}} away from GR;
this means expanding the fields as power series in a small parameter,
namely $\bar\zeta = (\bar\lambda/\lambda_{\GW})^{2}$. As in Eq.~\eqref{eq:gen-alt-h-exp-old}, we impose
\begin{equation}
\label{eq:gen-alt-h-exp}
\barh_{\mu\nu} = \sum_{n=0}^\infty \bar\zeta^n \barh^{(n)}_{\mu\nu}\,,
\end{equation}
and similarly for other fields, where the zeroth field
$\barh^{(0)}_{\mu\nu}$ is the GR solution. Inserting this expansion
in the first-order equation of motion
Eq.~\eqref{eq:rank2EOM1atscri} and matching order by order gives
\begin{subequations}
\label{eq:matchedorders}
\begin{eqnarray}
\label{eq:matchedordersn}
\kappa \bar\square\barh^{(n+1)}_{\mu\nu} &=&  - \left( 2\alpha_1 + \alpha_2 + 2 \alpha_3 \right) f(\bar\vartheta)
\nn \\
&&\qquad\times  \left( \bcd_\mu\bcd_\nu\bar\square - \met_{\mu\nu}
 \bar\square\bar\square \right) \barh^{(n)} \nn\\
&&\qquad {}- \left( \alpha_2 + 4\alpha_3
\right) f(\bar\vartheta) \bar\square\bar\square\barh^{(n)}_{\mu\nu} \,,
\end{eqnarray}
for all orders $n\ge 0$, and
\begin{equation}
\label{eq:matchedorders0}
\bar\square\barh^{(0)}_{\mu\nu} = 0 \,,
\end{equation}
for the GR solution. Substituting Eq.~\eqref{eq:matchedorders0} into
Eq.~\eqref{eq:matchedordersn} and iteratively solving the field
equations one order at a time, we find at all orders that
\begin{equation}
\label{eq:matchedordersall}
\bar\square\barh^{(n)}_{\mu\nu} = 0 \,.
\end{equation}
\end{subequations}
This is the GR first-order equation of motion, and just as in GR, we
may specialize the Lorenz gauge to the TT gauge at
$\scri^+$\footnote{To prove that the TT gauge exists at $\scri^+$ for
  this theory, the proof in Appendix~A of Flanagan and Hughes
  (FH)~\cite{Flanagan:2005p} must be extended. Their~Eq.~(A.12) must
  be replaced by our~\eqref{eq:rank2EOM1tensatscri} and a small
  coupling expansion performed. The result will again
  be~\eqref{eq:matchedordersall}, which is identical to
  FH's~Eq.~(A.12).}. This expansion has discontinuously turned the
massive wave equation into a massless one by killing the massive modes
in the limit of $m\to\infty$. Such an order-reduction procedure, where
certain solutions are eliminated through perturbative constraints, has
been shown to select the physically correct ones in all studied
cases~\cite{1996PhRvD..54.6233F,2007LNP...720..403W,2009PhRvD..79h4043Y}.

We can now evaluate the complete effective stress-energy tensor of GWs at $\scri^+$. 
As shown in Sec.~\ref{sec:gen-alt-dyn}, there is no direct contribution from the dynamical 
part of the interaction term. Section~\ref{sec:gen-alt-non-dyn} showed that the non-dynamical 
part does contribute directly, but imposing Eq.~\eqref{eq:matchedordersall} forces 
this contribution to also vanish. Since the MacCallum-Taub tensor on-shell is equal to the 
Isaacson tensor, we then have
\begin{equation}
\label{eq:gen-ESETonshell}
T^\eff_{\mu\nu} = \left(T^\eff_{\MT\mu\nu} + T^\eff_{\mbox{\tiny
      n-d}\mu\nu}\right)\Big|_{(\bar\square\barh_{\alpha\beta}=0)}
= T^\eff_{\GR\mu\nu} \,.
\end{equation}
That is, the effective GW stress-energy tensor is identical to the Isaacson one at $\scri^{+}$
for this wide class of modified gravity theories.

\section{Conclusions}
\label{sec:conc}
We have here addressed the energy content of GWs in a wide class of modified gravity theories.
We focused on theories that are weak deformations away from GR and calculated the effective
stress-energy tensor where GWs are extracted: in the asymptotically-flat region of spacetime.

The main calculation tool we employed was the perturbed Lagrangian
approach. We demonstrated the calculation explicitly for GR, 
recovering the Isaacson effective stress-energy tensor. We
also explicitly calculated this effective tensor in dynamical modified CS gravity, 
where again the result at $\scri^+$ reduces to the Isaacson tensor. The features of
CS gravity that lead to the effective stress-energy tensor being
identical to the one in GR are the dynamical nature of the scalar field and the topological
nature of the curvature correction to the action. 

We then generalized this finding to all action modifications of a
similar nature: a dynamical scalar field coupled to a scalar curvature
invariant of rank 2 or higher in a spacetime that is asymptotically
flat. For scalar curvature invariants of rank 3 or higher, we showed
that there is no modification to the stress-energy tensor or the
equations of motion at $\scri^+$. For rank 2, we calculated the
contribution to the effective stress-energy tensor and to the first-order 
equation of motion. In the weak coupling limit, the only
solutions to the first-order equations of motion satisfy
the GR first-order equations of motion at $\scri^+$, namely
$\bar\square\barh_{\mu\nu} = 0$. Evaluating the effective
stress-energy tensor on-shell with these solutions leads, again, to
the Isaacson stress-energy tensor.

A few caveats are in order. As we have stressed before, this result is
evaluated at asymptotically-flat, future null infinity, so it does not
apply to cosmological spacetimes, e.g.~de~Sitter spacetime. Not all of the
energy that is lost by a system is carried away by GWs to
$\scri^+$: there is also radiation in the scalar field (which is
calculated straightforwardly from $T_{\mu\nu}^{(\vartheta)}$), and
both GWs and the scalar field radiation are lost to trapped
surfaces. All of these effects must be accounted for in calculating
the radiation-reaction of a system. Finally, we did not address
modifications to the action of the form $f(\vartheta) R$, which reduce to
a classical scalar-tensor theory.

There are several avenues open for future work. Considering classical
scalar-tensor modifications is one possible extension. The work
should also be extended to the next simplest spacetimes, those 
that are asymptotically de~Sitter. This is appropriate for
calculating GWs from inflation, for example. Extending this approach
to calculating energy lost to trapped surfaces is another
possibility.

The most natural application of this work is in tests of GR with pulsar binaries
and with GWs emitted by EMRIs. The former problem
requires performing a post-Keplerian expansion of the motion of bodies
orbiting each other. The latter requires knowing the BH
spacetime (background) solution in the class of modified gravity theories and the 
geodesic or non-geodesic motion on that spacetime. Both of these programs require
knowledge of radiation-reaction in GWs at $\scri^+$,
which we have here computed for a large class of modified gravity
theories.

\begin{acknowledgments}
  We are grateful to Stephen Green, Ted Jacobson, Eanna Flanagan, Eric
  Poisson, Leor Barack, Nathan Johnson-McDaniel, Scott Hughes, and an
  anonymous referee for valuable discussions.  LCS acknowledges
  support from NSF Grant PHY-0449884 and the Solomon Buchsbaum Fund at
  MIT.  NY acknowledges support from the NASA through Einstein
  Postdoctoral Fellowship Award Number PF9-00063 and PF0-110080 issued
  by the Chandra X-ray Observatory Center, which is operated by the
  Smithsonian Astrophysical Observatory for and on behalf of the
  National Aeronautics Space Administration under contract NAS8-03060.
\end{acknowledgments}

\bibliography{master,phyjabb}

\begin{thebibliography}{10}%
\makeatletter
\providecommand \@ifxundefined [1]{%
 \ifx #1\undefined \expandafter \@firstoftwo
 \else \expandafter \@secondoftwo
\fi
}%
\providecommand \@ifnum [1]{%
 \ifnum #1\expandafter \@firstoftwo
 \else \expandafter \@secondoftwo
\fi
}%
\providecommand \enquote [1]{``#1''}%
\providecommand \bibnamefont  [1]{#1}%
\providecommand \bibfnamefont [1]{#1}%
\providecommand \citenamefont [1]{#1}%
\providecommand\href[0]{\@sanitize\@href}%
\providecommand\@href[1]{\endgroup\@@startlink{#1}\endgroup\@@href}%
\providecommand\@@href[1]{#1\@@endlink}%
\providecommand \@sanitize [0]{\begingroup\catcode`\&12\catcode`\#12\relax}%
\@ifxundefined \pdfoutput {\@firstoftwo}{%
 \@ifnum{\z@=\pdfoutput}{\@firstoftwo}{\@secondoftwo}%
}{%
 \providecommand\@@startlink[1]{\leavevmode\special{html:<a href="#1">}}%
 \providecommand\@@endlink[0]{\special{html:</a>}}%
}{%
 \providecommand\@@startlink[1]{%
  \leavevmode
  \pdfstartlink
   attr{/Border[0 0 1 ]/H/I/C[0 1 1]}%
   user{/Subtype/Link/A<</Type/Action/S/URI/URI(#1)>>}%
  \relax
 }%
 \providecommand\@@endlink[0]{\pdfendlink}%
}%
\providecommand \url  [0]{\begingroup\@sanitize \@url }%
\providecommand \@url [1]{\endgroup\@href {#1}{\urlprefix}}%
\providecommand \urlprefix [0]{URL }%
\providecommand \Eprint[0]{\href }%
\@ifxundefined \urlstyle {%
  \providecommand \doi [1]{doi:\discretionary{}{}{}#1}%
}{%
  \providecommand \doi [0]{doi:\discretionary{}{}{}\begingroup
  \urlstyle{rm}\Url }%
}%
\providecommand \doibase [0]{http://dx.doi.org/}%
\providecommand \Doi[1]{\href{\doibase#1}}%
\providecommand \bibAnnote [3]{%
  \BibitemShut{#1}%
  \begin{quotation}\noindent
    \textsc{Key:}\ #2\\\textsc{Annotation:}\ #3%
  \end{quotation}%
}%
\providecommand \bibAnnoteFile [2]{%
  \IfFileExists{#2}{\bibAnnote {#1} {#2} {\input{#2}}}{}%
}%
\providecommand \typeout [0]{\immediate \write \m@ne }%
\providecommand \selectlanguage [0]{\@gobble}%
\providecommand \bibinfo [0]{\@secondoftwo}%
\providecommand \bibfield [0]{\@secondoftwo}%
\providecommand \translation [1]{[#1]}%
\providecommand \BibitemOpen[0]{}%
\providecommand \bibitemStop [0]{}%
\providecommand \bibitemNoStop [0]{.\EOS\space}%
\providecommand \EOS [0]{\spacefactor3000\relax}%
\providecommand \BibitemShut [1]{\csname bibitem#1\endcsname}%
\bibitem{lrr-2006-3}%
  \BibitemOpen
  \bibfield{author}{%
  \bibinfo {author} {\bibfnamefont{C.~M.}\ \bibnamefont{Will}},\ }%
  \bibfield{journal}{%
  \bibinfo {journal} {Living Reviews in Relativity}\ }%
  \textbf{\bibinfo {volume} {9}} (\bibinfo {year} {2006}),\
  \Eprint{http://arxiv.org/abs/gr-qc/0510072}{arXiv:gr-qc/0510072},\
  \url{http://www.livingreviews.org/lrr-2006-3}%
  \bibAnnoteFile{NoStop}{lrr-2006-3}%
\bibitem{Burgay:2003jj}%
  \BibitemOpen
  \bibfield{author}{%
  \bibinfo {author} {\bibfnamefont{M.}~\bibnamefont{Burgay}} \emph{et~al.},\ }%
  \bibfield{journal}{%
  \Doi{10.1038/nature02124}{\bibinfo {journal} {Nature.}}\ }%
  \textbf{\bibinfo {volume} {426}},\ \bibinfo {pages} {531} (\bibinfo {year}
  {2003}),\
  \Eprint{http://arxiv.org/abs/astro-ph/0312071}{arXiv:astro-ph/0312071}%
  \bibAnnoteFile{NoStop}{Burgay:2003jj}%
\bibitem{Lyne:2004cj}%
  \BibitemOpen
  \bibfield{author}{%
  \bibinfo {author} {\bibfnamefont{A.~G.}\ \bibnamefont{Lyne}} \emph{et~al.},\
  }%
  \bibfield{journal}{%
  \Doi{10.1126/science.1094645}{\bibinfo {journal} {Science}}\ }%
  \textbf{\bibinfo {volume} {303}},\ \bibinfo {pages} {1153} (\bibinfo {year}
  {2004}),\
  \Eprint{http://arxiv.org/abs/astro-ph/0401086}{arXiv:astro-ph/0401086}%
  \bibAnnoteFile{NoStop}{Lyne:2004cj}%
\bibitem{Kramer:2006nb}%
  \BibitemOpen
  \bibfield{author}{%
  \bibinfo {author} {\bibfnamefont{M.}~\bibnamefont{Kramer}} \emph{et~al.},\ }%
  \bibfield{journal}{%
  \Doi{10.1126/science.1132305}{\bibinfo {journal} {Science}}\ }%
  \textbf{\bibinfo {volume} {314}},\ \bibinfo {pages} {97} (\bibinfo {year}
  {2006}),\
  \Eprint{http://arxiv.org/abs/astro-ph/0609417}{arXiv:astro-ph/0609417}%
  \bibAnnoteFile{NoStop}{Kramer:2006nb}%
\bibitem{Yunes:2008ua}%
  \BibitemOpen
  \bibfield{author}{%
  \bibinfo {author} {\bibfnamefont{N.}~\bibnamefont{Yunes}}\ and\ \bibinfo
  {author} {\bibfnamefont{D.~N.}\ \bibnamefont{Spergel}},\ }%
  \bibfield{journal}{%
  \Doi{10.1103/PhysRevD.80.042004}{\bibinfo {journal} {Phys. Rev.}}\ }%
  \textbf{\bibinfo {volume} {D80}},\ \bibinfo {pages} {042004} (\bibinfo {year}
  {2009}),\ \Eprint{http://arxiv.org/abs/0810.5541}{arXiv:0810.5541 [gr-qc]}%
  \bibAnnoteFile{NoStop}{Yunes:2008ua}%
\bibitem{2010PhRvD..82h2002Y}%
  \BibitemOpen
  \bibfield{author}{%
  \bibinfo {author} {\bibfnamefont{N.}~\bibnamefont{{Yunes}}}\ and\ \bibinfo
  {author} {\bibfnamefont{S.~A.}\ \bibnamefont{{Hughes}}},\ }%
  \bibfield{journal}{%
  \Doi{10.1103/PhysRevD.82.082002}{\bibinfo {journal} {\prd}}\ }%
  \textbf{\bibinfo {volume} {82}},\ \bibinfo {pages} {082002} (\bibinfo {month}
  {Oct.}\ \bibinfo {year} {2010}),\
  \Eprint{http://arxiv.org/abs/1007.1995}{arXiv:1007.1995 [gr-qc]}%
  \bibAnnoteFile{NoStop}{2010PhRvD..82h2002Y}%
\bibitem{ligo}%
  \BibitemOpen
  \enquote{\bibinfo {title} {{LIGO}},}\ \bibinfo {note} {{\tt
  www.ligo.caltech.edu}}%
  \bibAnnoteFile{NoStop}{ligo}%
\bibitem{Abramovici:1992ah}%
  \BibitemOpen
  \bibfield{author}{%
  \bibinfo {author} {\bibfnamefont{A.}~\bibnamefont{Abramovici}}
  \emph{et~al.},\ }%
  \bibfield{journal}{%
  \bibinfo {journal} {Science}\ }%
  \textbf{\bibinfo {volume} {256}},\ \bibinfo {pages} {325} (\bibinfo {year}
  {1992})%
  \bibAnnoteFile{NoStop}{Abramovici:1992ah}%
\bibitem{:2007kva}%
  \BibitemOpen
  \bibfield{author}{%
  \bibinfo {author} {\bibfnamefont{L.~S.}\ \bibnamefont{Collaboration}}
  (\bibinfo {collaboration} {LIGO Scientific})}%
   (\bibinfo {year} {2007}),\
  \Eprint{http://arxiv.org/abs/0711.3041}{arXiv:0711.3041 [gr-qc]}%
  \bibAnnoteFile{NoStop}{:2007kva}%
\bibitem{virgo}%
  \BibitemOpen
  \enquote{\bibinfo {title} {{VIRGO}},}\ \bibinfo {note} {{\tt
  www.virgo.infn.it}}%
  \bibAnnoteFile{NoStop}{virgo}%
\bibitem{lisa}%
  \BibitemOpen
  \enquote{\bibinfo {title} {{LISA}},}\ \bibinfo {note} {{\tt
  www.esa.int/science/lisa}, {\tt lisa.jpl.nasa.gov}}%
  \bibAnnoteFile{NoStop}{lisa}%
\bibitem{Prince:2003aa}%
  \BibitemOpen
  \bibfield{author}{%
  \bibinfo {author} {\bibfnamefont{T.}~\bibnamefont{{Prince}}},\ }%
  \bibfield{journal}{%
  \bibinfo {journal} {American Astronomical Society Meeting}\ }%
  \textbf{\bibinfo {volume} {202}},\ \bibinfo {pages} {3701} (\bibinfo {month}
  {May}\ \bibinfo {year} {2003})%
  \bibAnnoteFile{NoStop}{Prince:2003aa}%
\bibitem{Danzmann:2003tv}%
  \BibitemOpen
  \bibfield{author}{%
  \bibinfo {author} {\bibfnamefont{K.}~\bibnamefont{Danzmann}}\ and\ \bibinfo
  {author} {\bibfnamefont{A.}~\bibnamefont{Rudiger}},\ }%
  \bibfield{journal}{%
  \bibinfo {journal} {Class. Quant. Grav.}\ }%
  \textbf{\bibinfo {volume} {20}},\ \bibinfo {pages} {S1} (\bibinfo {year}
  {2003})%
  \bibAnnoteFile{NoStop}{Danzmann:2003tv}%
\bibitem{Danzmann:2003ad}%
  \BibitemOpen
  \bibfield{author}{%
  \bibinfo {author} {\bibfnamefont{K.}~\bibnamefont{{Danzmann}}},\ }%
  \bibfield{journal}{%
  \bibinfo {journal} {Advances in Space Research}\ }%
  \textbf{\bibinfo {volume} {32}},\ \bibinfo {pages} {1233} (\bibinfo {month}
  {Oct.}\ \bibinfo {year} {2003})%
  \bibAnnoteFile{NoStop}{Danzmann:2003ad}%
\bibitem{Schutz:2009tz}%
  \BibitemOpen
  \bibfield{author}{%
  \bibinfo {author} {\bibfnamefont{B.~F.}\ \bibnamefont{Schutz}}, \bibinfo
  {author} {\bibfnamefont{J.}~\bibnamefont{Centrella}}, \bibinfo {author}
  {\bibfnamefont{C.}~\bibnamefont{Cutler}},\ and\ \bibinfo {author}
  {\bibfnamefont{S.~A.}\ \bibnamefont{Hughes}}}%
   (\bibinfo {year} {2009}),\
  \Eprint{http://arxiv.org/abs/0903.0100}{arXiv:0903.0100 [gr-qc]}%
  \bibAnnoteFile{NoStop}{Schutz:2009tz}%
\bibitem{Gralla:2010cd}%
  \BibitemOpen
  \bibfield{author}{%
  \bibinfo {author} {\bibfnamefont{S.~E.}\ \bibnamefont{Gralla}},\ }%
  \bibfield{journal}{%
  \Doi{10.1103/PhysRevD.81.084060}{\bibinfo {journal} {Phys. Rev.}}\ }%
  \textbf{\bibinfo {volume} {D81}},\ \bibinfo {pages} {084060} (\bibinfo {year}
  {2010}),\ \Eprint{http://arxiv.org/abs/1002.5045}{arXiv:1002.5045 [gr-qc]}%
  \bibAnnoteFile{NoStop}{Gralla:2010cd}%
\bibitem{Alexander:2009tp}%
  \BibitemOpen
  \bibfield{author}{%
  \bibinfo {author} {\bibfnamefont{S.}~\bibnamefont{Alexander}}\ and\ \bibinfo
  {author} {\bibfnamefont{N.}~\bibnamefont{Yunes}},\ }%
  \bibfield{journal}{%
  \Doi{10.1016/j.physrep.2009.07.002}{\bibinfo {journal} {Phys. Rept.}}\ }%
  \textbf{\bibinfo {volume} {480}},\ \bibinfo {pages} {1} (\bibinfo {year}
  {2009}),\ \Eprint{http://arxiv.org/abs/0907.2562}{arXiv:0907.2562 [hep-th]}%
  \bibAnnoteFile{NoStop}{Alexander:2009tp}%
\bibitem{Barack:2009ux}%
  \BibitemOpen
  \bibfield{author}{%
  \bibinfo {author} {\bibfnamefont{L.}~\bibnamefont{Barack}},\ }%
  \bibfield{journal}{%
  \Doi{10.1088/0264-9381/26/21/213001}{\bibinfo {journal} {Class. Quant.
  Grav.}}\ }%
  \textbf{\bibinfo {volume} {26}},\ \bibinfo {pages} {213001} (\bibinfo {year}
  {2009}),\ \Eprint{http://arxiv.org/abs/0908.1664}{arXiv:0908.1664 [gr-qc]}%
  \bibAnnoteFile{NoStop}{Barack:2009ux}%
\bibitem{Pound:2007th}%
  \BibitemOpen
  \bibfield{author}{%
  \bibinfo {author} {\bibfnamefont{A.}~\bibnamefont{Pound}}\ and\ \bibinfo
  {author} {\bibfnamefont{E.}~\bibnamefont{Poisson}},\ }%
  \bibfield{journal}{%
  \Doi{10.1103/PhysRevD.77.044013}{\bibinfo {journal} {Phys. Rev.}}\ }%
  \textbf{\bibinfo {volume} {D77}},\ \bibinfo {pages} {044013} (\bibinfo {year}
  {2008}),\ \Eprint{http://arxiv.org/abs/0708.3033}{arXiv:0708.3033 [gr-qc]}%
  \bibAnnoteFile{NoStop}{Pound:2007th}%
\bibitem{Gralla:2008fg}%
  \BibitemOpen
  \bibfield{author}{%
  \bibinfo {author} {\bibfnamefont{S.~E.}\ \bibnamefont{Gralla}}\ and\ \bibinfo
  {author} {\bibfnamefont{R.~M.}\ \bibnamefont{Wald}},\ }%
  \bibfield{journal}{%
  \Doi{10.1088/0264-9381/25/20/205009}{\bibinfo {journal} {Class. Quant.
  Grav.}}\ }%
  \textbf{\bibinfo {volume} {25}},\ \bibinfo {pages} {205009} (\bibinfo {year}
  {2008}),\ \Eprint{http://arxiv.org/abs/0806.3293}{arXiv:0806.3293 [gr-qc]}%
  \bibAnnoteFile{NoStop}{Gralla:2008fg}%
\bibitem{Pound:2009sm}%
  \BibitemOpen
  \bibfield{author}{%
  \bibinfo {author} {\bibfnamefont{A.}~\bibnamefont{Pound}},\ }%
  \bibfield{journal}{%
  \Doi{10.1103/PhysRevD.81.024023}{\bibinfo {journal} {Phys. Rev.}}\ }%
  \textbf{\bibinfo {volume} {D81}},\ \bibinfo {pages} {024023} (\bibinfo {year}
  {2010}),\ \Eprint{http://arxiv.org/abs/0907.5197}{arXiv:0907.5197 [gr-qc]}%
  \bibAnnoteFile{NoStop}{Pound:2009sm}%
\bibitem{Pound:2010pj}%
  \BibitemOpen
  \bibfield{author}{%
  \bibinfo {author} {\bibfnamefont{A.}~\bibnamefont{Pound}},\ }%
  \bibfield{journal}{%
  \Doi{10.1103/PhysRevD.81.124009}{\bibinfo {journal} {Phys. Rev.}}\ }%
  \textbf{\bibinfo {volume} {D81}},\ \bibinfo {pages} {124009} (\bibinfo {year}
  {2010}),\ \Eprint{http://arxiv.org/abs/1003.3954}{arXiv:1003.3954 [gr-qc]}%
  \bibAnnoteFile{NoStop}{Pound:2010pj}%
\bibitem{Pound:2010wa}%
  \BibitemOpen
  \bibfield{author}{%
  \bibinfo {author} {\bibfnamefont{A.}~\bibnamefont{Pound}}}%
   (\bibinfo {year} {2010}),\
  \Eprint{http://arxiv.org/abs/1006.3903}{arXiv:1006.3903 [gr-qc]}%
  \bibAnnoteFile{NoStop}{Pound:2010wa}%
\bibitem{Isaacson:1968ra}%
  \BibitemOpen
  \bibfield{author}{%
  \bibinfo {author} {\bibfnamefont{R.~A.}\ \bibnamefont{{Isaacson}}},\ }%
  \bibfield{journal}{%
  \Doi{10.1103/PhysRev.166.1263}{\bibinfo {journal} {Phys. Rev.}}\ }%
  \textbf{\bibinfo {volume} {166}},\ \bibinfo {pages} {1263} (\bibinfo {month}
  {Feb.}\ \bibinfo {year} {1968})%
  \bibAnnoteFile{NoStop}{Isaacson:1968ra}%
\bibitem{Isaacson:1968gw}%
  \BibitemOpen
  \bibfield{author}{%
  \bibinfo {author} {\bibfnamefont{R.~A.}\ \bibnamefont{{Isaacson}}},\ }%
  \bibfield{journal}{%
  \Doi{10.1103/PhysRev.166.1272}{\bibinfo {journal} {Phys. Rev.}}\ }%
  \textbf{\bibinfo {volume} {166}},\ \bibinfo {pages} {1272} (\bibinfo {month}
  {Feb.}\ \bibinfo {year} {1968})%
  \bibAnnoteFile{NoStop}{Isaacson:1968gw}%
\bibitem{Nelson:2010rt}%
  \BibitemOpen
  \bibfield{author}{%
  \bibinfo {author} {\bibfnamefont{W.}~\bibnamefont{Nelson}}, \bibinfo {author}
  {\bibfnamefont{J.}~\bibnamefont{Ochoa}},\ and\ \bibinfo {author}
  {\bibfnamefont{M.}~\bibnamefont{Sakellariadou}},\ }%
  \bibfield{journal}{%
  \Doi{10.1103/PhysRevD.82.085021}{\bibinfo {journal} {Phys. Rev.}}\ }%
  \textbf{\bibinfo {volume} {D82}},\ \bibinfo {pages} {085021} (\bibinfo {year}
  {2010}),\ \Eprint{http://arxiv.org/abs/1005.4276}{arXiv:1005.4276 [hep-th]}%
  \bibAnnoteFile{NoStop}{Nelson:2010rt}%
\bibitem{Nelson:2010ru}%
  \BibitemOpen
  \bibfield{author}{%
  \bibinfo {author} {\bibfnamefont{W.}~\bibnamefont{Nelson}}, \bibinfo {author}
  {\bibfnamefont{J.}~\bibnamefont{Ochoa}},\ and\ \bibinfo {author}
  {\bibfnamefont{M.}~\bibnamefont{Sakellariadou}},\ }%
  \bibfield{journal}{%
  \Doi{10.1103/PhysRevLett.105.101602}{\bibinfo {journal} {Phys. Rev. Lett.}}\
  }%
  \textbf{\bibinfo {volume} {105}},\ \bibinfo {pages} {101602} (\bibinfo {year}
  {2010}),\ \Eprint{http://arxiv.org/abs/1005.4279}{arXiv:1005.4279 [hep-th]}%
  \bibAnnoteFile{NoStop}{Nelson:2010ru}%
\bibitem{Sopuerta:2009iy}%
  \BibitemOpen
  \bibfield{author}{%
  \bibinfo {author} {\bibfnamefont{C.~F.}\ \bibnamefont{Sopuerta}}\ and\
  \bibinfo {author} {\bibfnamefont{N.}~\bibnamefont{Yunes}},\ }%
  \bibfield{journal}{%
  \Doi{10.1103/PhysRevD.80.064006}{\bibinfo {journal} {Phys. Rev.}}\ }%
  \textbf{\bibinfo {volume} {D80}},\ \bibinfo {pages} {064006} (\bibinfo {year}
  {2009}),\ \Eprint{http://arxiv.org/abs/0904.4501}{arXiv:0904.4501 [gr-qc]}%
  \bibAnnoteFile{NoStop}{Sopuerta:2009iy}%
\bibitem{Guarrera:2007tu}%
  \BibitemOpen
  \bibfield{author}{%
  \bibinfo {author} {\bibfnamefont{D.}~\bibnamefont{Guarrera}}\ and\ \bibinfo
  {author} {\bibfnamefont{A.~J.}\ \bibnamefont{Hariton}}}%
   (\bibinfo {year} {2007}),\
  \Eprint{http://arxiv.org/abs/gr-qc/0702029}{gr-qc/0702029}%
  \bibAnnoteFile{NoStop}{Guarrera:2007tu}%
\bibitem{Grumiller:2007rv}%
  \BibitemOpen
  \bibfield{author}{%
  \bibinfo {author} {\bibfnamefont{D.}~\bibnamefont{Grumiller}}\ and\ \bibinfo
  {author} {\bibfnamefont{N.}~\bibnamefont{Yunes}},\ }%
  \bibfield{journal}{%
  \Doi{10.1103/PhysRevD.77.044015}{\bibinfo {journal} {Phys. Rev.}}\ }%
  \textbf{\bibinfo {volume} {D77}},\ \bibinfo {pages} {044015} (\bibinfo {year}
  {2008}),\ \Eprint{http://arxiv.org/abs/0711.1868}{arXiv:0711.1868 [gr-qc]}%
  \bibAnnoteFile{NoStop}{Grumiller:2007rv}%
\bibitem{2009PhRvD..79h4043Y}%
  \BibitemOpen
  \bibfield{author}{%
  \bibinfo {author} {\bibfnamefont{N.}~\bibnamefont{{Yunes}}}\ and\ \bibinfo
  {author} {\bibfnamefont{F.}~\bibnamefont{{Pretorius}}},\ }%
  \bibfield{journal}{%
  \Doi{10.1103/PhysRevD.79.084043}{\bibinfo {journal} {\prd}}\ }%
  \textbf{\bibinfo {volume} {79}},\ \bibinfo {pages} {084043} (\bibinfo {month}
  {Apr.}\ \bibinfo {year} {2009}),\
  \Eprint{http://arxiv.org/abs/0902.4669}{arXiv:0902.4669 [gr-qc]}%
  \bibAnnoteFile{NoStop}{2009PhRvD..79h4043Y}%
\bibitem{jackiw:2003:cmo}%
  \BibitemOpen
  \bibfield{author}{%
  \bibinfo {author} {\bibfnamefont{R.}~\bibnamefont{Jackiw}}\ and\ \bibinfo
  {author} {\bibfnamefont{S.~Y.}\ \bibnamefont{Pi}},\ }%
  \bibfield{journal}{%
  \bibinfo {journal} {Phys. Rev.}\ }%
  \textbf{\bibinfo {volume} {D68}},\ \bibinfo {pages} {104012} (\bibinfo {year}
  {2003}),\ \Eprint{http://arxiv.org/abs/gr-qc/0308071}{gr-qc/0308071}%
  \bibAnnoteFile{NoStop}{jackiw:2003:cmo}%
\bibitem{Alexander:2007kv}%
  \BibitemOpen
  \bibfield{author}{%
  \bibinfo {author} {\bibfnamefont{S.}~\bibnamefont{Alexander}}, \bibinfo
  {author} {\bibfnamefont{L.~S.}\ \bibnamefont{Finn}},\ and\ \bibinfo {author}
  {\bibfnamefont{N.}~\bibnamefont{Yunes}},\ }%
  \bibfield{journal}{%
  \Doi{10.1103/PhysRevD.78.066005}{\bibinfo {journal} {Phys. Rev.}}\ }%
  \textbf{\bibinfo {volume} {D78}},\ \bibinfo {pages} {066005} (\bibinfo {year}
  {2008}),\ \Eprint{http://arxiv.org/abs/0712.2542}{arXiv:0712.2542 [gr-qc]}%
  \bibAnnoteFile{NoStop}{Alexander:2007kv}%
\bibitem{Yunes:2010yf}%
  \BibitemOpen
  \bibfield{author}{%
  \bibinfo {author} {\bibfnamefont{N.}~\bibnamefont{Yunes}}, \bibinfo {author}
  {\bibfnamefont{R.}~\bibnamefont{O'Shaughnessy}}, \bibinfo {author}
  {\bibfnamefont{B.~J.}\ \bibnamefont{Owen}},\ and\ \bibinfo {author}
  {\bibfnamefont{S.}~\bibnamefont{Alexander}},\ }%
  \bibfield{journal}{%
  \Doi{10.1103/PhysRevD.82.064017}{\bibinfo {journal} {Phys. Rev.}}\ }%
  \textbf{\bibinfo {volume} {D82}},\ \bibinfo {pages} {064017} (\bibinfo {year}
  {2010}),\ \Eprint{http://arxiv.org/abs/1005.3310}{arXiv:1005.3310 [gr-qc]}%
  \bibAnnoteFile{NoStop}{Yunes:2010yf}%
\bibitem{2005PThPh.113..481N}%
  \BibitemOpen
  \bibfield{author}{%
  \bibinfo {author} {\bibfnamefont{K.}~\bibnamefont{{Nakamura}}},\ }%
  \bibfield{journal}{%
  \Doi{10.1143/PTP.113.481}{\bibinfo {journal} {Progress of Theoretical
  Physics}}\ }%
  \textbf{\bibinfo {volume} {113}},\ \bibinfo {pages} {481} (\bibinfo {month}
  {Mar.}\ \bibinfo {year} {2005}),\
  \Eprint{http://arxiv.org/abs/arXiv:gr-qc/0410024}{arXiv:gr-qc/0410024}%
  \bibAnnoteFile{NoStop}{2005PThPh.113..481N}%
\bibitem{2007arXiv0711.0996N}%
  \BibitemOpen
  \bibfield{author}{%
  \bibinfo {author} {\bibfnamefont{K.}~\bibnamefont{{Nakamura}}},\ }%
  \bibfield{journal}{%
  \bibinfo {journal} {ArXiv e-prints}}%
   (\bibinfo {month} {Nov.}\ \bibinfo {year} {2007}),\
  \Eprint{http://arxiv.org/abs/0711.0996}{arXiv:0711.0996 [gr-qc]}%
  \bibAnnoteFile{NoStop}{2007arXiv0711.0996N}%
\bibitem{1964PhRv..135..271B}%
  \BibitemOpen
  \bibfield{author}{%
  \bibinfo {author} {\bibfnamefont{D.~R.}\ \bibnamefont{{Brill}}}\ and\
  \bibinfo {author} {\bibfnamefont{J.~B.}\ \bibnamefont{{Hartle}}},\ }%
  \bibfield{journal}{%
  \Doi{10.1103/PhysRev.135.B271}{\bibinfo {journal} {Physical Review}}\ }%
  \textbf{\bibinfo {volume} {135}},\ \bibinfo {pages} {271} (\bibinfo {month}
  {Jul.}\ \bibinfo {year} {1964})%
  \bibAnnoteFile{NoStop}{1964PhRv..135..271B}%
\bibitem{1992GReGr..24.1015Z}%
  \BibitemOpen
  \bibfield{author}{%
  \bibinfo {author} {\bibfnamefont{R.~M.}\ \bibnamefont{{Zalaletdinov}}},\ }%
  \bibfield{journal}{%
  \Doi{10.1007/BF00756944}{\bibinfo {journal} {General Relativity and
  Gravitation}}\ }%
  \textbf{\bibinfo {volume} {24}},\ \bibinfo {pages} {1015} (\bibinfo {month}
  {Oct.}\ \bibinfo {year} {1992})%
  \bibAnnoteFile{NoStop}{1992GReGr..24.1015Z}%
\bibitem{1996GReGr..28..953Z}%
  \BibitemOpen
  \bibfield{author}{%
  \bibinfo {author} {\bibfnamefont{R.~M.}\ \bibnamefont{{Zalaletdinov}}},\ }%
  \bibfield{journal}{%
  \Doi{10.1007/BF02113091}{\bibinfo {journal} {General Relativity and
  Gravitation}}\ }%
  \textbf{\bibinfo {volume} {28}},\ \bibinfo {pages} {953} (\bibinfo {month}
  {Aug.}\ \bibinfo {year} {1996})%
  \bibAnnoteFile{NoStop}{1996GReGr..28..953Z}%
\bibitem{2004gr.qc....11004Z}%
  \BibitemOpen
  \bibfield{author}{%
  \bibinfo {author} {\bibfnamefont{R.}~\bibnamefont{{Zalaletdinov}}},\ }%
  \bibfield{journal}{%
  \bibinfo {journal} {ArXiv General Relativity and Quantum Cosmology
  e-prints}}%
   (\bibinfo {month} {Oct.}\ \bibinfo {year} {2004}),\
  \Eprint{http://arxiv.org/abs/arXiv:gr-qc/0411004}{arXiv:gr-qc/0411004}%
  \bibAnnoteFile{NoStop}{2004gr.qc....11004Z}%
\bibitem{Misner:1973cw}%
  \BibitemOpen
  \bibfield{author}{%
  \bibinfo {author} {\bibfnamefont{C.~W.}\ \bibnamefont{Misner}}, \bibinfo
  {author} {\bibfnamefont{K.}~\bibnamefont{Thorne}},\ and\ \bibinfo {author}
  {\bibfnamefont{J.~A.}\ \bibnamefont{Wheeler}},\ }%
  \emph{\bibinfo {title} {Gravitation}}\ (\bibinfo {publisher} {W. H. Freeman
  \& Co.},\ \bibinfo {address} {San Francisco},\ \bibinfo {year} {1973})%
  \bibAnnoteFile{NoStop}{Misner:1973cw}%
\bibitem{Carroll}%
  \BibitemOpen
  \bibfield{author}{%
  \bibinfo {author} {\bibfnamefont{S.~M.}\ \bibnamefont{{Carroll}}},\ }%
  \emph{\bibinfo {title} {{Spacetime and geometry: An introduction to general
  relativity}}}\ (\bibinfo {publisher} {Benjamin Cummings},\ \bibinfo {year}
  {2004})%
  \bibAnnoteFile{NoStop}{Carroll}%
\bibitem{1973CMaPh..30..153M}%
  \BibitemOpen
  \bibfield{author}{%
  \bibinfo {author} {\bibfnamefont{M.~A.~H.}\ \bibnamefont{{MacCallum}}}\ and\
  \bibinfo {author} {\bibfnamefont{A.~H.}\ \bibnamefont{{Taub}}},\ }%
  \bibfield{journal}{%
  \Doi{10.1007/BF01645977}{\bibinfo {journal} {Communications in Mathematical
  Physics}}\ }%
  \textbf{\bibinfo {volume} {30}},\ \bibinfo {pages} {153} (\bibinfo {month}
  {Jun.}\ \bibinfo {year} {1973})%
  \bibAnnoteFile{NoStop}{1973CMaPh..30..153M}%
\bibitem{Yunes:2009ef}%
  \BibitemOpen
  \bibfield{author}{%
  \bibinfo {author} {\bibfnamefont{N.}~\bibnamefont{Yunes}}, \bibinfo {author}
  {\bibfnamefont{A.}~\bibnamefont{Buonanno}}, \bibinfo {author}
  {\bibfnamefont{S.~A.}\ \bibnamefont{Hughes}}, \bibinfo {author}
  {\bibfnamefont{M.}~\bibnamefont{Coleman~Miller}},\ and\ \bibinfo {author}
  {\bibfnamefont{Y.}~\bibnamefont{Pan}},\ }%
  \bibfield{journal}{%
  \Doi{10.1103/PhysRevLett.104.091102}{\bibinfo {journal} {Phys. Rev. Lett.}}\
  }%
  \textbf{\bibinfo {volume} {104}},\ \bibinfo {pages} {091102} (\bibinfo {year}
  {2010}),\ \Eprint{http://arxiv.org/abs/0909.4263}{arXiv:0909.4263 [gr-qc]}%
  \bibAnnoteFile{NoStop}{Yunes:2009ef}%
\bibitem{Yunes:2010zj}%
  \BibitemOpen
  \bibfield{author}{%
  \bibinfo {author} {\bibfnamefont{N.}~\bibnamefont{Yunes}} \emph{et~al.}}%
   (\bibinfo {year} {2010}),\
  \Eprint{http://arxiv.org/abs/1009.6013}{arXiv:1009.6013 [gr-qc]}%
  \bibAnnoteFile{NoStop}{Yunes:2010zj}%
\bibitem{Blanchet:2002av}%
  \BibitemOpen
  \bibfield{author}{%
  \bibinfo {author} {\bibfnamefont{L.}~\bibnamefont{Blanchet}},\ }%
  \bibfield{journal}{%
  \bibinfo {journal} {Living Rev. Relativity}\ }%
  \textbf{\bibinfo {volume} {9}},\ \bibinfo {pages} {4} (\bibinfo {year}
  {2006}),\ \Eprint{http://arxiv.org/abs/gr-qc/0202016}{gr-qc/0202016}%
  \bibAnnoteFile{NoStop}{Blanchet:2002av}%
\bibitem{Mino:1997p287}%
  \BibitemOpen
  \bibfield{author}{%
  \bibinfo {author} {\bibfnamefont{Y.}~\bibnamefont{{Mino}}}, \bibinfo {author}
  {\bibfnamefont{M.}~\bibnamefont{{Sasaki}}}, \bibinfo {author}
  {\bibfnamefont{M.}~\bibnamefont{{Shibata}}}, \bibinfo {author}
  {\bibfnamefont{H.}~\bibnamefont{{Tagoshi}}},\ and\ \bibinfo {author}
  {\bibfnamefont{T.}~\bibnamefont{{Tanaka}}},\ }%
  \bibfield{journal}{%
  \Doi{10.1143/PTPS.128.1}{\bibinfo {journal} {Progress of Theoretical Physics
  Supplement}}\ }%
  \textbf{\bibinfo {volume} {128}},\ \bibinfo {pages} {1} (\bibinfo {year}
  {1997}),\
  \Eprint{http://arxiv.org/abs/arXiv:gr-qc/9712057}{arXiv:gr-qc/9712057}%
  \bibAnnoteFile{NoStop}{Mino:1997p287}%
\bibitem{Cambiaso:2010un}%
  \BibitemOpen
  \bibfield{author}{%
  \bibinfo {author} {\bibfnamefont{M.}~\bibnamefont{Cambiaso}}\ and\ \bibinfo
  {author} {\bibfnamefont{L.~F.}\ \bibnamefont{Urrutia}}}%
   (\bibinfo {year} {2010}),\
  \Eprint{http://arxiv.org/abs/1010.4526}{arXiv:1010.4526 [gr-qc]}%
  \bibAnnoteFile{NoStop}{Cambiaso:2010un}%
\bibitem{2007CoPhC.177..640M}%
  \BibitemOpen
  \bibfield{author}{%
  \bibinfo {author}
  {\bibfnamefont{J.}~\bibnamefont{{Mart{\'\i}n-Garc{\'\i}a}}}, \bibinfo
  {author} {\bibfnamefont{R.}~\bibnamefont{{Portugal}}},\ and\ \bibinfo
  {author} {\bibfnamefont{L.}~\bibnamefont{{Manssur}}},\ }%
  \bibfield{journal}{%
  \Doi{10.1016/j.cpc.2007.05.015}{\bibinfo {journal} {Computer Physics
  Communications}}\ }%
  \textbf{\bibinfo {volume} {177}},\ \bibinfo {pages} {640} (\bibinfo {month}
  {Oct.}\ \bibinfo {year} {2007}),\
  \Eprint{http://arxiv.org/abs/0704.1756}{arXiv:0704.1756 [cs.SC]}%
  \bibAnnoteFile{NoStop}{2007CoPhC.177..640M}%
\bibitem{2008CoPhC.179..586M}%
  \BibitemOpen
  \bibfield{author}{%
  \bibinfo {author} {\bibfnamefont{J.~M.}\
  \bibnamefont{{Mart{\'{\i}}n-Garc{\'{\i}}a}}}, \bibinfo {author}
  {\bibfnamefont{D.}~\bibnamefont{{Yllanes}}},\ and\ \bibinfo {author}
  {\bibfnamefont{R.}~\bibnamefont{{Portugal}}},\ }%
  \bibfield{journal}{%
  \Doi{10.1016/j.cpc.2008.04.018}{\bibinfo {journal} {Computer Physics
  Communications}}\ }%
  \textbf{\bibinfo {volume} {179}},\ \bibinfo {pages} {586} (\bibinfo {month}
  {Oct.}\ \bibinfo {year} {2008}),\
  \Eprint{http://arxiv.org/abs/0802.1274}{arXiv:0802.1274 [cs.SC]}%
  \bibAnnoteFile{NoStop}{2008CoPhC.179..586M}%
\bibitem{2008CoPhC.179..597M}%
  \BibitemOpen
  \bibfield{author}{%
  \bibinfo {author} {\bibfnamefont{J.~M.}\
  \bibnamefont{{Mart{\'{\i}}n-Garc{\'{\i}}a}}},\ }%
  \bibfield{journal}{%
  \Doi{10.1016/j.cpc.2008.05.009}{\bibinfo {journal} {Computer Physics
  Communications}}\ }%
  \textbf{\bibinfo {volume} {179}},\ \bibinfo {pages} {597} (\bibinfo {month}
  {Oct.}\ \bibinfo {year} {2008}),\
  \Eprint{http://arxiv.org/abs/0803.0862}{arXiv:0803.0862 [cs.SC]}%
  \bibAnnoteFile{NoStop}{2008CoPhC.179..597M}%
\bibitem{2009GReGr..41.2415B}%
  \BibitemOpen
  \bibfield{author}{%
  \bibinfo {author} {\bibfnamefont{D.}~\bibnamefont{{Brizuela}}}, \bibinfo
  {author} {\bibfnamefont{J.~M.}\
  \bibnamefont{{Mart{\'{\i}}n-Garc{\'{\i}}a}}},\ and\ \bibinfo {author}
  {\bibfnamefont{G.~A.}\ \bibnamefont{{Mena Marug{\'a}n}}},\ }%
  \bibfield{journal}{%
  \Doi{10.1007/s10714-009-0773-2}{\bibinfo {journal} {General Relativity and
  Gravitation}}\ }%
  \textbf{\bibinfo {volume} {41}},\ \bibinfo {pages} {2415} (\bibinfo {month}
  {Oct.}\ \bibinfo {year} {2009}),\
  \Eprint{http://arxiv.org/abs/0807.0824}{arXiv:0807.0824 [gr-qc]}%
  \bibAnnoteFile{NoStop}{2009GReGr..41.2415B}%
\bibitem{Xact}%
  \BibitemOpen
  \enquote{\bibinfo {title} {{xAct}},}\ \bibinfo {note} {{\tt
  http://www.xact.es/}}%
  \bibAnnoteFile{NoStop}{Xact}%
\bibitem{Alexander:2004wk}%
  \BibitemOpen
  \bibfield{author}{%
  \bibinfo {author} {\bibfnamefont{S.}~\bibnamefont{Alexander}}\ and\ \bibinfo
  {author} {\bibfnamefont{J.}~\bibnamefont{Martin}},\ }%
  \bibfield{journal}{%
  \bibinfo {journal} {Phys. Rev.}\ }%
  \textbf{\bibinfo {volume} {D71}},\ \bibinfo {pages} {063526} (\bibinfo {year}
  {2005}),\ \Eprint{http://arxiv.org/abs/hep-th/0410230}{hep-th/0410230}%
  \bibAnnoteFile{NoStop}{Alexander:2004wk}%
\bibitem{2010PhRvD..82f4017Y}%
  \BibitemOpen
  \bibfield{author}{%
  \bibinfo {author} {\bibfnamefont{N.}~\bibnamefont{{Yunes}}}, \bibinfo
  {author} {\bibfnamefont{R.}~\bibnamefont{{O'Shaughnessy}}}, \bibinfo {author}
  {\bibfnamefont{B.~J.}\ \bibnamefont{{Owen}}},\ and\ \bibinfo {author}
  {\bibfnamefont{S.}~\bibnamefont{{Alexander}}},\ }%
  \bibfield{journal}{%
  \Doi{10.1103/PhysRevD.82.064017}{\bibinfo {journal} {\prd}}\ }%
  \textbf{\bibinfo {volume} {82}},\ \bibinfo {pages} {064017} (\bibinfo {month}
  {Sep.}\ \bibinfo {year} {2010}),\
  \Eprint{http://arxiv.org/abs/1005.3310}{arXiv:1005.3310 [gr-qc]}%
  \bibAnnoteFile{NoStop}{2010PhRvD..82f4017Y}%
\bibitem{Thorne:1980rm}%
  \BibitemOpen
  \bibfield{author}{%
  \bibinfo {author} {\bibfnamefont{K.~S.}\ \bibnamefont{Thorne}},\ }%
  \bibfield{journal}{%
  \bibinfo {journal} {Rev. Mod. Phys.}\ }%
  \textbf{\bibinfo {volume} {52}},\ \bibinfo {pages} {299} (\bibinfo {year}
  {1980})%
  \bibAnnoteFile{NoStop}{Thorne:1980rm}%
\bibitem{Blanchet:2005rj}%
  \BibitemOpen
  \bibfield{author}{%
  \bibinfo {author} {\bibfnamefont{L.}~\bibnamefont{Blanchet}}, \bibinfo
  {author} {\bibfnamefont{M.~S.~S.}\ \bibnamefont{Qusailah}},\ and\ \bibinfo
  {author} {\bibfnamefont{C.~M.}\ \bibnamefont{Will}},\ }%
  \bibfield{journal}{%
  \bibinfo {journal} {\apj}\ }%
  \textbf{\bibinfo {volume} {635}},\ \bibinfo {pages} {508} (\bibinfo {year}
  {2005}),\ \Eprint{http://arxiv.org/abs/astro-ph/0507692}{astro-ph/0507692}%
  \bibAnnoteFile{NoStop}{Blanchet:2005rj}%
\bibitem{Polchinski:1998rr}%
  \BibitemOpen
  \bibfield{author}{%
  \bibinfo {author} {\bibfnamefont{J.}~\bibnamefont{Polchinski}},\ }%
  \emph{\bibinfo {title} {String theory. Vol. 2: Superstring theory and
  beyond}}\ (\bibinfo {publisher} {Cambridge University Press},\ \bibinfo
  {address} {Cambridge, UK},\ \bibinfo {year} {1998})%
  \bibAnnoteFile{NoStop}{Polchinski:1998rr}%
\bibitem{Yunes:2007ss}%
  \BibitemOpen
  \bibfield{author}{%
  \bibinfo {author} {\bibfnamefont{N.}~\bibnamefont{Yunes}}\ and\ \bibinfo
  {author} {\bibfnamefont{C.~F.}\ \bibnamefont{Sopuerta}},\ }%
  \bibfield{journal}{%
  \Doi{10.1103/PhysRevD.77.064007}{\bibinfo {journal} {Phys. Rev.}}\ }%
  \textbf{\bibinfo {volume} {D77}},\ \bibinfo {pages} {064007} (\bibinfo {year}
  {2008}),\ \Eprint{http://arxiv.org/abs/0712.1028}{arXiv:0712.1028 [gr-qc]}%
  \bibAnnoteFile{NoStop}{Yunes:2007ss}%
\bibitem{1993tegp.book.....W}%
  \BibitemOpen
  \bibfield{author}{%
  \bibinfo {author} {\bibfnamefont{C.~M.}\ \bibnamefont{{Will}}},\ }%
  \emph{\bibinfo {title} {Theory and Experiment in Gravitational Physics, by
  Clifford M.~Will, pp.~396.~ISBN 0521439736.~Cambridge, UK: Cambridge
  University Press, March 1993.}}\ (\bibinfo {year} {1993})%
  \bibAnnoteFile{NoStop}{1993tegp.book.....W}%
\bibitem{alexander:2004:lfg}%
  \BibitemOpen
  \bibfield{author}{%
  \bibinfo {author} {\bibfnamefont{S.~H.~S.}\ \bibnamefont{Alexander}},
  \bibinfo {author} {\bibfnamefont{M.~E.}\ \bibnamefont{Peskin}},\ and\
  \bibinfo {author} {\bibfnamefont{M.~M.}\ \bibnamefont{Sheikh-Jabbari}},\ }%
  \bibfield{journal}{%
  \bibinfo {journal} {Phys. Rev. Lett.}\ }%
  \textbf{\bibinfo {volume} {96}},\ \bibinfo {pages} {081301} (\bibinfo {year}
  {2006}),\ \Eprint{http://arxiv.org/abs/hep-th/0403069}{hep-th/0403069}%
  \bibAnnoteFile{NoStop}{alexander:2004:lfg}%
\bibitem{Alexander:2004xd}%
  \BibitemOpen
  \bibfield{author}{%
  \bibinfo {author} {\bibfnamefont{S.~H.~S.}\ \bibnamefont{Alexander}}\ and\
  \bibinfo {author} {\bibfnamefont{J.}~\bibnamefont{Gates},
  \bibfnamefont{S.~James}},\ }%
  \bibfield{journal}{%
  \bibinfo {journal} {JCAP}\ }%
  \textbf{\bibinfo {volume} {0606}},\ \bibinfo {pages} {018} (\bibinfo {year}
  {2006}),\ \Eprint{http://arxiv.org/abs/hep-th/0409014}{hep-th/0409014}%
  \bibAnnoteFile{NoStop}{Alexander:2004xd}%
\bibitem{2009PhRvD..80f5003G}%
  \BibitemOpen
  \bibfield{author}{%
  \bibinfo {author} {\bibfnamefont{S.~J.}\ \bibnamefont{{Gates}},
  \bibfnamefont{Jr.}}, \bibinfo {author} {\bibfnamefont{S.~V.}\
  \bibnamefont{{Ketov}}},\ and\ \bibinfo {author}
  {\bibfnamefont{N.}~\bibnamefont{{Yunes}}},\ }%
  \bibfield{journal}{%
  \Doi{10.1103/PhysRevD.80.065003}{\bibinfo {journal} {\prd}}\ }%
  \textbf{\bibinfo {volume} {80}},\ \bibinfo {pages} {065003} (\bibinfo {month}
  {Sep.}\ \bibinfo {year} {2009}),\
  \Eprint{http://arxiv.org/abs/0906.4978}{arXiv:0906.4978 [hep-th]}%
  \bibAnnoteFile{NoStop}{2009PhRvD..80f5003G}%
\bibitem{Flanagan:2005p}%
  \BibitemOpen
  \bibfield{author}{%
  \bibinfo {author} {\bibfnamefont{{\'E}.~{\'E}.}\ \bibnamefont{{Flanagan}}}\
  and\ \bibinfo {author} {\bibfnamefont{S.~A.}\ \bibnamefont{{Hughes}}},\ }%
  \bibfield{journal}{%
  \Doi{10.1088/1367-2630/7/1/204}{\bibinfo {journal} {New Journal of Physics}}\
  }%
  \textbf{\bibinfo {volume} {7}},\ \bibinfo {pages} {204} (\bibinfo {month}
  {Sep.}\ \bibinfo {year} {2005}),\
  \Eprint{http://arxiv.org/abs/arXiv:gr-qc/0501041}{arXiv:gr-qc/0501041}%
  \bibAnnoteFile{NoStop}{Flanagan:2005p}%
\bibitem{1996PhRvD..54.6233F}%
  \BibitemOpen
  \bibfield{author}{%
  \bibinfo {author} {\bibfnamefont{{\'E}.~{\'E}.}\ \bibnamefont{{Flanagan}}}\
  and\ \bibinfo {author} {\bibfnamefont{R.~M.}\ \bibnamefont{{Wald}}},\ }%
  \bibfield{journal}{%
  \Doi{10.1103/PhysRevD.54.6233}{\bibinfo {journal} {\prd}}\ }%
  \textbf{\bibinfo {volume} {54}},\ \bibinfo {pages} {6233} (\bibinfo {month}
  {Nov.}\ \bibinfo {year} {1996}),\
  \Eprint{http://arxiv.org/abs/arXiv:gr-qc/9602052}{arXiv:gr-qc/9602052}%
  \bibAnnoteFile{NoStop}{1996PhRvD..54.6233F}%
\bibitem{2007LNP...720..403W}%
  \BibitemOpen
  \bibfield{author}{%
  \bibinfo {author} {\bibfnamefont{R.}~\bibnamefont{{Woodard}}},\ }%
  in\ \emph{\bibinfo {booktitle} {The Invisible Universe: Dark Matter and Dark
  Energy}},\ \bibinfo {series} {Lecture Notes in Physics, Berlin Springer
  Verlag}, Vol.\ \bibinfo {volume} {720},\ \bibinfo {editor} {edited by\
  \bibinfo {editor} {\bibnamefont{{L.~Papantonopoulos}}}}\ (\bibinfo {year}
  {2007})\ pp.\ \bibinfo {pages} {403--+},\
  \Eprint{http://arxiv.org/abs/arXiv:astro-ph/0601672}{arXiv:astro-ph/0601672}%
  \bibAnnoteFile{NoStop}{2007LNP...720..403W}%
\end{thebibliography}%
\end{document}